\documentclass[useAMS,usenatbib,twocolumn]{mn2e}
\usepackage{amsmath}
\usepackage{graphicx}
\usepackage{color}
\usepackage{epstopdf}

\newcommand{\be}{\begin{equation}}
\newcommand{\ee}{\end{equation}}
\newcommand{\bea}{\begin{eqnarray}}
\newcommand{\eea}{\end{eqnarray}}
\newcommand{\bfig}{\begin{figure}}
\newcommand{\efig}{\end{figure}}
\newcommand{\bfigs}{\begin{figure*}}
\newcommand{\efigs}{\end{figure*}}
\newcommand{\bt}{\begin{table}}
\newcommand{\et}{\end{table}}

\def\simlt{\lower.5ex\hbox{$\; \buildrel < \over \sim \;$}}
\def\simgt{\lower.5ex\hbox{$\; \buildrel > \over \sim \;$}}
\def\simpt{\lower.5ex\hbox{$\; \buildrel \propto \over \sim \;$}}

\def\kms{\mbox{km s$^{-1}$}}

\def\pc{\mbox{pc}}
\def\Da{\mathcal{D}}

\definecolor{mylabelcolor}{rgb}{0.5,1,1}

\title{GLAMER Part I: A Code for Gravitational Lensing Simulations with Adaptive Mesh Refinement}

\date{\today}

\author[Metcalf \& Petkova]{ R. Benton Metcalf$^1$ \& Margarita Petkova$^{1,2,3}$
  \\ \\ $^{1}$ Dipartimento di Fisica e
  Astronomia, Universit\`a di Bologna, viale Berti Pichat 6/2,
  40127, Bologna, Italy \\
  $^{2}$  Department of Physics, Ludwig-Maximilians-Universit\"at, Scheinerstr. 1, D-81679 M\"unchen, Germany\\
  $^{3}$  Excellence Cluster Universe, Boltzmannstr. 2, D-85748 Garching, Germany
}
\begin{document}

\maketitle

\begin{abstract}
A computer code is described for the simulation of gravitational lensing data.  The code incorporates adaptive mesh refinement in choosing which rays to shoot based on the requirements of the source size, location and surface brightness distribution or to find critical curves/caustics.  A variety of source surface brightness models are implemented to represent galaxies and quasar emission regions.  The lensing mass can be represented by point masses (stars), smoothed simulation particles, analytic halo models, pixelized mass maps or any combination of these.  The deflection and beam distortions (convergence and shear) are calculated by modified tree algorithm when halos, point masses or particles are used and by FFT when mass maps are used.  The combination of these methods allow for a very large dynamical range to be represented in a single simulation.  Individual images of galaxies can be represented in a simulation that covers many square degrees.  For an individual strongly lensed quasar, source sizes from the size of the quasar's host galaxy ($\sim 100$~kpc) down to microlensing scales ($\sim 10^{-4}$~pc) can be probed in a self consistent simulation.  Descriptions of various tests of the code's accuracy are given.
\end{abstract}

\begin{keywords}
\end{keywords}

\section{introduction}

Gravitational lensing is becoming a more and more powerful tool for investigating cosmology and astrophysics. It has been used to study such disparate subjects as dark energy, the Cold Dark Matter (CDM) model for structure formation, the formation and structure of galaxies, the structure of active galactic nuclei (AGN) and the distribution of stars and planets in and around our own galaxy.  In most cases, detailed comparisons between theoretical models and lensing data require computer simulations.


Here we seek to develop a computer code, called GLAMER, flexible enough that it can be used to simulate many different types of lensing data include weak lensing on large scales from cosmological simulations, weak and strong lensing by galaxy clusters, galaxy-galaxy lensing, galaxy-quasar lensing, lensing by substructures in the lenses or somewhere along the line-of-sight and microlensing by stars in a lens galaxy.  To achieve this the code needs to be able to represent the sources and mass distribution within the lens in a very flexible and configurable way.  

One of our goals is to study microlensing in the same simulation as strong lensing on galaxy scales.   A galaxy acting as a strong lens for a QSO might have a scale of $\sim 100$~kpc while the microlensing of the quasar by stars might take place on a scale of $\sim 0.01$~kpc or smaller.  From its host galaxy down to its X-ray emitting accretion disk, the QSO has emission regions on all scales that can be used to simultaneously constrain the lens and source model.  A related goal is to simulate the images of galaxies in the same simulation as weak lensing across many square degrees.  This would enable one to use different redshifts for each source and to represent the effects of lensing on all relevant scales simultaneously.  Both these goals require a very large dynamical range in resolution for the ray-shooting. 

 
 Another problem we wish to address with GLAMER is that of the inner regions of strong lenses typically being dominated by baryons -- stars and gas -- which are not represented in the highest resolution N-body simulations.  It is also true that lensing quantities such as the magnification are very sensitive to the numerical noise that comes from the discreteness of the simulation particles.  Recently \cite{2013arXiv1309.1161A} have made some progress on this latter problem.  Our approach is to allow for a combination of smoothed particles and analytic profiles to exist within the simulation.


We also wanted to make the code flexible enough to do many different lensing applications in a relatively user friendly way.  To this end our code allows for a variety of source types -- circular, analytic galaxy models, pixelized images, multiple sources, QSO emission regions, etc. -- which can be easily added to.  It also allows for a variety of lenses -- analytic halos, analytic galaxies, N-body particles, stars, smooth shear fields  --- which can also be easily added to.  The code also has the ability to do multiplane lensing although this capability will be discussed in a companion paper \citep{PM&G2013}.  Here we will concentrate on what has been implemented to allow this flexibility and dynamic range.

There is an extensive literature on simulating gravitational lensing
\cite[for a very sparse sample:][]{
1998ApJ...494...29W,
1999MNRAS.306..567F,
2000ApJ...530..547J,
 2001MNRAS.327..169H,
AMCO04,
2003ApJ...592..699V,
2006ApJ...653..942M,
2007A&A...471..731P,
2007MNRAS.382..121H,
2007MNRAS.376..113A,
2008A&A...482..403M,
2009ApJ...701..945S,
2009A&A...499...31H,
2010ApJ...712..658P,
2010MNRAS.tmp.1510V,
2011ApJ...742...15T,
2012MNRAS.420..155K,
2013arXiv1309.1161A}.  GLAMER makes significant advances beyond previous codes in its dynamical range, adaptability to a wide variety of applications and ease of use.  We will refer to some previous work as it becomes relevant to the discussion.
 
In section~\ref{sec:lensing-simulations}, the basic problem that the code is required to solve is described.  The calculation of the deflection angle is discussed in section~\ref{sec:calc-defl}.  The methods used for adaptive mesh refinement, image finding and caustic finding are discussed in section~\ref{sec:find-mapp-imag}.   In section~\ref{sec:tests}, various tests of the code are presented and the paper is concluded in section~\ref{sec:discussion}.

\section{lensing simulations}
\label{sec:lensing-simulations}

The job of a ray-tracing code can be broken up into two broad tasks.  One is to calculate the deflection angle which will be discussed in section~\ref{sec:calc-defl} and the other is to find and map the images which will be discussed in section~\ref{sec:find-mapp-imag}.  At the heart of these tasks is the lensing equation which relates a point on the source-plane, ${\bf y}$, to its image point, {\bf x}, on the image-plane:
\begin{equation}\label{lens_eq}
{\bf y} = {\bf x} - \boldsymbol{\alpha}({\bf x}).
\end{equation}

In this paper, we will restrict ourselves to a single lens plane.  The more complicated case of multiple lens planes will be discussed in a companion paper~\citep{PM&G2013}.  For a single lens plane, the coordinates can be related to points on the sky by
\begin{align}
{\bf x} = \Da_l \boldsymbol{\theta} ~~~~,~~~~ {\bf y} = \Da_l \boldsymbol{\beta} 
\end{align}
where $\Da_l$ is a reference distance which will usually be taken to be the angular size distance to the lens in the case of a single lens-plane.  The deflection angle, $\boldsymbol{\alpha}({\bf x})$, (which has units of length in this form) can be related to the true deflection angle in the path of a light-ray, $\tilde{\boldsymbol{\alpha}}({\bf x})$, if the lens consists of a single plane by
\begin{align}
\boldsymbol{\alpha}({\bf x}) = \frac{\Da_l\Da_{ls}}{\Da_s}~ \tilde{\boldsymbol{\alpha}}({\bf x})
\end{align}
where $\Da_s$ is the angular size distance to the source and $\Da_{ls}$ is the angular size distance between the lens and the source.  In the case of multiple lens planes, $\boldsymbol{\alpha}({\bf x})$ can be interpreted as simply the displacement of the light-ray when it reaches the source plane projected onto a plane at $\Da_l$.  Another very useful quantity is the critical surface density defined as
\begin{align}
\Sigma_{\rm crit} \equiv \frac{c^2}{4\pi G} \frac{\Da_s}{\Da_l\Da_{ls}}
\end{align}
where $G$ is Newton's constant and $c$ is the speed of light.

\section{calculating the deflection}
\label{sec:calc-defl}

GLAMER has several options for calculating the deflection angle, $\pmb\alpha({\bf x})$, designed for different applications.  The lens can be represented by an analytic model in which case the surface density, deflection and shear at each point ${\bf x}$ can be calculated analytically. Analytic models are discussed at length elsewhere (see for example \cite{SEF92}) and the implementation in GLAMER code is not greatly different so they will not be discussed further here.  
 A more difficult case arises when the lens consists of many components.  These might be particles from a N-body/SPH simulation, individual stars in the lensing galaxy, or many clumps and subclumps of matter represented by analytic profiles.  In these cases, it might not be optimal or practical to simply sum up the contributions from each component because of their large number, perhaps billions.  In the case of N-body/SPH simulations, there is the additional complication of smoothing the mass distribution.

\subsection{smoothing}
\label{sec:smoothing}

A method for smoothing the density distribution is required when dealing with N-body/SPH particles from a simulation.  GLAMER uses an SPH--like approach where the
smoothing scale for a particle, $r_{\rm sm}$, is the radius around the particle
which contains the nearest $N_{\rm sph}$ particles.  This gives an adaptive
smoothing which is smaller where the particles are denser.  

The question arises as to whether to smooth according to the number of particle neighbors in three dimensions or in projection, two dimensions.  The former is clearly truer to the intent of the simulation, but the latter is faster.
\cite{2006ApJ...652...43L} looked at several different smoothing methods in lensing simulations and found that smoothing in 3 dimensions is better than in 2.  Our experiments confirm this conclusion.  Two dimensional smoothing tends to under smooth the low density particles in the foreground and background of the lens. These particles often constitute a substantial fraction if not the majority of the total integrated density along the line-of-sight.  Note that in the 3D case there is not a unique smoothing scale for a point on the image plane which introduces an additional complication that will be discussed in section~\ref{sec:calc-lens-quant}.  The 3D smoothing can be done once for a simulation since it is valid for all projections.   

Calculating the number of particle neighbors is facilitated by sorting
the particles into a tree structure which is also used in the
deflection calculation that is described in the next section.  This procedure is parallelizable in a straightforward way and need only be done once per simulation so it is not a significant bottleneck.

The mass of each particle is treated as if it were distributed as a 2D
Gaussian 
\begin{align}\label{eq:smoothing}
\Sigma({\bf x}) = \frac{m}{2\pi r_{\rm sm}^2} \frac{ \exp\left[- \frac{|{\bf x}|^2}{2 r_{\rm sm}^2} \right]}{\left[ 1 - e^{-q^2/2} \right]} 
\times \left\{ 
\begin{array}{ccl}
1 & , & |{\bf x}| < q \, r_{\rm sm}\\
0& , & |{\bf x}| > q \, r_{\rm sm}
\end{array}
\right.
\end{align}
where $m$ is the particle mass and $q$ is a cutoff so that it has compact support.  Other profiles are possible, for example the traditional SPH polynomial kernel, but (\ref{eq:smoothing}) is what is usually used in GLAMER.

In some cases, the simulation has particles of different masses.  These may represent different components - stars, gas, dark matter - or the structure under consideration might have been re-simulated with higher resolution so that there are small mass particles near the center and very massive particles that represent the larger-scale structure surrounding the lens.  In this case, each species of particle is considered separately, the smoothing scale depends only on the density of particles of that species.  If this is not done the density distribution can become unrealistically grainy; very massive particles on the outskirts of a high resolution region can be given very small smoothing lengths because of their many less massive neighbor particles.


\subsection{calculating the lensing quantities} 
\label{sec:calc-lens-quant}

The lensing equation (\ref{lens_eq}) relates points on the lens--plane, ${\bf x}$, to points
on the source--plane, ${\bf y}$.
The deflection angle $\boldsymbol{\alpha}({\bf x})$ and
the derivatives of the lens equation need to be calculated.  By tradition the derivatives are grouped into the
convergence, $\kappa({\bf x})$, two components of shear, $\boldsymbol{\gamma}({\bf x})$ 
and time delay, $\delta t({\bf x})$ , which are defined as
\begin{align} \label{alpha}
&\boldsymbol{\alpha}({\bf x}) = \boldsymbol{\nabla} \psi({\bf x})
\\ \label{kappa}
&\kappa({\bf x}) = \frac{1}{2} {\rm tr}\left[ \frac{\partial \boldsymbol{\alpha}}{\partial
    {\bf x} } \right] 
 = \frac{\Sigma({\bf x})}{\Sigma_{\rm crit}}
\\ \label{gamma_1}
&\gamma_1({\bf x}) = \frac{1}{2} \left[ \frac{\partial  \alpha_1}{\partial
     x_1} - \frac{\partial \alpha_2}{\partial
    x_2 }  \right] \\ \label{gamma_2}
&\gamma_2({\bf x}) = \frac{\partial  \alpha_2}{\partial x_1}
 = \frac{\partial  \alpha_1}{\partial x_2} \\ \label{timedelay}
&\delta t({\bf x}) = \frac{1}{2} \left| \boldsymbol{\alpha} ( {\bf x}) \right|^2 
- \psi ( {\bf x})
\end{align}
where the second equality in (\ref{kappa}) and \eqref{timedelay} are valid only for a single, thin lens plane.  
The potential can be calculated from the Poisson equation
\begin{align}\label{eq:poisson_eq}
\nabla^2 \psi({\bf x})= 2\, \kappa({\bf x}).
\end{align}
The magnification of a point is $\mu({\bf x}) = \left[ (1-\kappa)^2 -  |\boldsymbol{\gamma}|^2 \right]^{-1}$.
Quantities (\ref{alpha}) through (\ref{timedelay}) are the lensing quantities
that we wish to calculate.   However, when calculating the images of finite sized sources, it is often the case that only the deflection angle (\ref{alpha}) is needed.  For this reason the deflection solver has an option that turns off the calculation of these quantities to save time.

The majority of ray-shooting simulations in the past have used a Fast Fourier Transform (FFT) based deflection solver.  In this scheme, the surface density is interpolated onto a regular, 2D grid \citep[for example][]{AMCO04,2008A&A...482..403M,2009A&A...499...31H}.  The lensing quantities can then be found by FFT through their connection to the lensing potential, \eqref{alpha} through \eqref{eq:poisson_eq}.  This method has the advantage of being fast --
scaling as $N \log N$ and not $N^2$ as in the case of a direct summation.  The FFT method, however, does suffer from several
drawbacks. The first is that the resolution of the lensing calculation
is limited to the grid size on which the FFT is performed.  This introduces an additional scale into the problem.  When working with simulation particles, the conservative thing to do is to make the grid scale much smaller than the smoothing scale everywhere on the plane.  This can be very inefficient in regions where the smoothing scale is large.  It is possible to use a multi-grid approach \cite{hilbert2007} where a smaller region around the area of interest is gridded at higher resolution.  This adds complexity especially when combined with some adaptive scheme as will be described in section~\ref{sec:find-mapp-imag}.  Also, 
the FFT method often requires padding the field with zeros to neutralize  periodic boundary conditions.  This further expands the grid.  The time bottleneck with the FFT is usually interpolating the mass to the grid.  This makes it laborious when many projections of the density or many changes in the mass distribution are needed when lens fitting for example.  Since the entire high resolution grid needs to be stored, the FFT approach can require a great deal of memory especially when many lens planes are needed (see Paper II).  Finally, point masses representing stars cannot be handled with a pure FFT approach.

GLAMER has the ability to read in mass maps and use them as lens planes.  In this case, it uses the FFT approach to find the lensing quantities.  In all other cases, for the reasons given above and others, GLAMER adopts a tree deflection solver very similar to tree force solvers commonly used for N-body simulations \citep{1989ApJS...70..389B}.  Some modifications have been made to the algorithm to improve its performance for lensing applications as will be discussed below.  This method starts with an expression for the lensing quantities for a single particle or halo.  It can be applied to both simulation particles and clumps of matter with analytic profiles which might represent galaxies, dark matter halos, subhalos, stars or other components of a lens model.  Here we will concentrate on simulation particles and make note of the places where other types of ``particles'' differ.
 
In the code, there are expressions for all of the lensing quantities
for each type of ``particle'' (or halo).
For example, the deflection angle for the smoothed particle~\eqref{eq:smoothing} is
\begin{align}
\boldsymbol{\alpha}&({\bf x}) = \frac{m}{\pi\Sigma_{\rm crit}} ~ \frac{{\bf x} }{|{\bf x}|^2} \\
&\times \left\{ 
\begin{array}{ccl}
 [ 1-e^{-\frac{|{\bf x}|^2}{2r_{\rm sm}}} ]/[ 1-e^{-q^2/2} ] & , & |{\bf x}| < q\, r_{\rm sm}\\
1 & , & |{\bf x}| > q \,r_{\rm sm}
\end{array}
\right.
\end{align}
\citep[see][]{2007MNRAS.376..113A}.
Outside of the outer radius of any spherically symmetric mass profile all the lensing quantities are identical to those for a point-mass.  In some cases, the analytic expressions contain special functions that are time consuming to evaluate (NFW halos for example).  In these cases a look-up table is created and the values are interpolated from the tables when they are needed.

All the lensing quantities obey the superposition principle so the contribution from each particle or halo can be added together to form the total.  The particles are sorted into a quad-tree structure.  This is a tree data structure where each of the branches contains four equal area sub-branches or children.  We have experimented with binary trees and equal particle number divisions of the boxes, but found this equal area quad-tree method to be superior.  A branch is not subdivided and becomes a {\it leaf} when it contains less than a certain number of particles, typically 5.


As the tree is constructed, the monopole and quadrupole moments of the mass in each box are calculated.  These moments are calculated with only the projected, 2D, distances of the particles.  Also while the tree is constructed, the quantity 
\begin{align}\label{r_cut}
&r_{\rm cut} \equiv 
  \frac{r_{\rm cm}}{\theta_{\rm force}} 
\end{align}
is calculated for each box where 
$r_{\rm cm}$ is the largest distance from the center of mass to an edge of the box.  The quantity $\theta_{\rm force}$ is an adjustable parameter that needs to be set after numerical convergence tests (see section~\ref{sec:tests}).  At the same time as the moments are calculated, a list is made of particles within that branch that have sizes larger than half the box size and have not been included in any parent branch's list.  The purpose of this list will be explained later.

To calculate the deflection at a point, or ray,  ${\bf x}$ the code ``walks the tree''.  As it reaches each box it evaluates the {\it multipole acceptability criterion} (MAC) which in this case is
\begin{align}
&|{\bf x} - {\bf x}_{\rm cm}| > r_{\rm cut} ~~~~~~~~ \mbox{in 2D}
\end{align}
where ${\bf x}_{\rm cm}$ is the box's center of mass.  If this expression is true the multipole moments of this box are used to calculate the lensing quantities.  If the expression is false the code descends into the box's children and repeats the process.  When the code reaches a leaf -- a box with no children -- the particles inside this box are added by direct summation.  In this way boxes that fill a small angle as ``seen'' from the point ${\bf x}$ are taken as a single unit because of (\ref{r_cut}).  

An additional problem arises in the lensing application of this algorithm that is not present for an N-body or SPH simulation.  This we will call the {\it big particle problem}.  In the lensing case, the 2D position of a particle might not be related to its size.  This might arise because small halos might be within larger halos or, in the case of smoothed N-body particles,  the size of the particles depends of the 3D position so background and foreground particles in low density regions will be much larger than the particles near the centers of the halos/galaxies.  An additional term can be added to the MAC, (\ref{r_cut}), containing the size of the biggest particle within the branch.  In this way direct summation would be triggered for any particle that the ray intersects.  This solution is highly inefficient in some circumstances.  The algorithm will always descend to the leafs containing the particles the ray intersects with and in the process will over resolve neighboring branches that may contain many particles that do not intersect the ray.  Our solution is to keep a list of particles in each branch that have sizes close to the size of the branch.  This allows for a fast method of identifying which particles intersect the ray so that they can be added individually and their contribution removed from the standard tree calculation.  This amounts to essentially pre-sorting the particles by 2D position and size so that they can be searched according to a combination of these.  We have experimented with other methods to solve the big particles problem, but have found this to be the most efficient.
The accuracy of this algorithm is tested by comparison to direct summation.

When the line-of-sight through a simulation is changed the particle positions are updated in place 
and a new tree structure is constructed with the z-axis
corresponding to the line of sight.  Having boxes aligned with the line of
sight rather than at an arbitrary angle significantly reduces the time spent
calculating the lensing quantities which more than makes up for the initial
cost of building the tree. 

There is a choice of whether to use a 3D or 2D tree in the calculation of the
lensing quantities.  With our solution to the big particle problem and our decision to use the planer lens approximation the 2D tree is more efficient in memory and speed although a 3D tree option is still implemented. 


\subsection{parallelization}
\label{sec:parallelization}
This tree deflection solving algorithm is easily parallelized by distributing the points on the image plane that need to be calculated amongst processors.  This becomes more efficient the bigger the chunks of grid points that are done in each call to the deflection solver, see section~\ref{sec:find-mapp-imag}.  Currently GLAMER uses POSIX threads to distribute the points into separate threads to be executed by different cores on a shared memory machine.

\subsection{Interpolation}
\label{sec:interpolation}

It often occurs that a grid cell needs to be refined based on a
criterion related to the source's surface brightness (see
section~\ref{sec:find-mapp-imag}), but the grid cell is small enough
compared to a structure in the lens that the lensing can be considered a linear distortion within the
cell.  Shooting rays for the newly added grid points with the full
deflection solver is inefficient.  GLAMER uses a scheme that
automatically detects if the magnification matrix is uniform over the
cell and uses a linear expansion of the deflection angle based on the
surrounding grid points to find the source positions of the new
subcells.  This method requires no a priori assumption about the scale
of structures in the lens relative to the source.  The process is
complicated by the fact that the grid is not uniform, but the
machinery developed for finding and mapping images and discussed in
section~\ref{sec:find-mapp-imag} make the basic required 
straightforward.

As will be discussed in section~\ref{sec:find-mapp-imag}, cells are defined by adding rays around an already shot ray.  When the magnification matrix is uniform across the cell the source
position of any point within the cell can be found from the position
of the center of the cell ${\bf x}_{\rm center}$ and its source
position ${\bf y}_{\rm center}$ which was already calculated in an
earlier stage of refinement:
\begin{equation} \label{eq:linear_lens_equation}
{\bf y} = {\bf y}_{\rm center} + {\mathcal A}_{\rm center} ({\bf x} - {\bf x}_{\rm center})
\end{equation}
where ${\mathcal A}_{\rm center}$ is the magnification matrix
(\ref{eq:magmatrix}) which can be found from the lensing quantities
\eqref{kappa} through \eqref{gamma_2}.

For each cell that is to be subdivided the code finds all of its
neighbor cells.   The $\kappa$ and ${\pmb \gamma}$ at these points are
compared to the center of the cell and if they are found to be within
a small tolerance (typically $\pm 10^{-4}$) the
approximation \eqref{eq:linear_lens_equation} is used. 
 
It is sometimes advantageous for precision and speed to not calculate 
$\kappa$ and ${\pmb \gamma}$.  Calculating these quantities for a
collection of asymmetric masses can incur a significant time penalty.
Also to save memory the $\kappa$ and ${\pmb \gamma}$ values at each
point are stored to floating point precision, but it is often
necessary to calculate the deflection to higher precision. 
 In these cases, the magnification matrix
can be found by using the deflection angles of the neighboring points
only.  Appendix~\ref{appendix:magn-matr-from} shows how this is done.
Sets of two neighbors are used to find ${\mathcal A}$ and if
they all agree to the tolerance level, the average ${\mathcal A}$ is
used to calculated the deflection in \eqref{eq:linear_lens_equation}.

\subsection{implanting stars in a smooth background}
\label{sec:impl-stars-smooth}

GLAMER has the ability to do lensing by a uniform distribution of
stars or by stars contained within a much larger lens galaxy.  The
deflection caused by stars can be calculated in the same way as for
any other ``particle'', but in this case the particle size is zero.
Stars can cause {\it microlensing} where what would have been a single
image is broken up into many micro-images which are generally not
individually resolved because of their small separation ($\sim$
micro-arcseconds).  The distinguishable separate patches of images are 
referred to as macro-images.

It is not possible, or necessary, to represent all of the
stars in a lens galaxy ($\sim 10^{11}$) as particles.  Only stars
relatively close to the images are needed.  The stars cannot simply be
added to the smooth component of the lens because the smooth mass
distribution is meant to represent the stars, gas and the dark matter
so the mass in stars must be subtracted from the smooth model near the
images.   The code must also take care not to overpopulate regions from
separate macro-images that might overlap.

When implanting the stars, the code subtracts a uniform
density disk equal to the mass in stars surrounding each macro-image.  
The positions of the macro-images are found by either taking
observed ones or finding them initially without stars. The stars are
then given random positions within each disk region.   In this way,
structure in the lensing on all scales from the size of the whole dark matter halo down
to tens a AU can be represented.

The deflection from each star is
\begin{equation}
\boldsymbol{\alpha}_*({\bf r}) = \frac{m}{\pi\Sigma_{\rm crit}} 
  ~ \frac{\hat{r}}{r} = \frac{r_{\rm E}^2}{r} ~ \hat{r} .
\end{equation}
where this defines the Einstein radius $r_{\rm E}$.  The same tree algorithm with a 2D tree is used in this case to calculate the total deflection.  Each of the stars may have equal masses or different masses randomly drawn from a mass function.

\cite{SEF92} estimate that the number of stars that need to be represented in a simulation is $\sim 100 \langle \mu \rangle \kappa^2_*$ where  $\langle \mu \rangle$ is the magnification if there were no stars and $\kappa_*$ is the surface density of stars in units of the critical density.  This criterion is always amply satisfied.  Typically 10,000 stars per macro-image are used.  Good convergence is found with this large number of stars.

\subsection{mass sheets}
\label{sec:mass-sheets}

As mentioned earlier, it is possible to read a map of the surface mass density into GLAMER and use it as a lens.  In this case the lensing quantities are calculated by FFT and interpolated from the grid.  This is particularly useful when dealing with N-body/SPH simulations of entire light cones on cosmological scales where the number of particles can be very large.  The particle distribution can be smoothed and projected onto planes which results in a very large savings in memory usage.

\subsection{multiple deflection planes}
\label{sec:mult-defl-plan}

GLAMER does have the ability to calculate the deflection from multiple
planes as is necessary for ray-tracing through a large-scale
cosmological N-body simulations.  This ability is discussed in detail in a
companion paper \citep{PM&G2013}. 

\section{finding and mapping the images}
\label{sec:find-mapp-imag}

Generally a source's position and shape is chosen and the task is to find its
images.  The number, size and shape of images are not known a priori.  Some images can be highly elongated which presents a problem when doing
calculations on a rectangular grid.  Some images can be much smaller than
others.  For finding the images of point sources the problem can be reduced to 
minimizing $|{\bf y}({\bf x}) - {\bf y}_{\rm source}|$ with respect to ${\bf x}$ where ${\bf y}_{\rm
  source}$ is the source position.  As in many minimization problems,
difficulties arise from not knowing the number of minima (images) and in
separating close images.   For finite size sources the most widely used 
method is to simply fill the image plane with a uniform grid, calculate the
source position of each point and find which points are within the source.
This is highly inefficient if ray-shooting is expensive since the grid spacing needs to be fine enough that
many points will land inside the smallest image that one needs to resolve.
For small sources with widely separated images the problem is especially
acute.  It is also important for cosmological simulations where the images of high redshift galaxies are typically shifted by arcminutes while their images can be smaller than an arcsecond.

We adopt an adaptive mesh refinement (AMR) scheme.  The points on the
source plane are linked to their image 
points and vice versa.   The source and image points are sorted into
separate kd-trees so that they can be quickly searched for nearest
neighbors and neighbor points within a fixed distance.  These kd-trees are ``live'' in the sense that points can be continuously added to them as the grid is refined.

We will first describe how a circular, uniform surface brightness source is found and refined, 
and then discuss sources with varying surface brightness.  Because lensing conserves surface brightness, in the case of uniform surface brightness the magnification is found by finding the area of the image and dividing by area of the source.  A point on the image-plane whose corresponding point on the source-plane is within the source is a point in the image.

The first step is to find the images.  On a coarse grid one would not expect to find any points within any image that is much smaller than the grid resolution.
To avoid this problem, we employ a telescoping source strategy.
First an initial coarse grid is put down with grid spacing $\Delta_{\rm
  init}$.  The source size is first set to 
$\Delta y_{\rm source} = \Delta_{\rm init}/\sqrt{\mu_{\rm min}}$ where
$\mu_{\rm min}$ is the absolute value of the magnification of 
the smallest image that one is required to find.  If the source is not
circular, $\Delta y_{\rm source}$ represents the largest linear dimension of
the source.  All the points within source are found.  If any of these
points have a grid size larger than $\Delta y_{\rm
  source}\sqrt{\mu_{\rm min}}/3$ it is refined.  A grid cell is always
refined by dividing it into nine subcells and thus reducing the grid
spacing by 1/3. After the refinements the source size is reduced by
1/3 and the process is repeated.  This is done until the desired source
size is reached at which point a more complex refinement scheme is
adopted.  Sometimes the full range of telescoping is not necessary when the
source has not moved, but has only changed in size.  It is possible for this algorithm to miss small, separated images while refining other images to a much smaller scale than the missed image.  The extent to which this happens is investigated in section~\ref{sec:imagefindingtests}.

When the target source size is reached in the telescoping stage the code separates individual images by a neighbors-of-neighbors algorithm on the image points.  Neighbors are cells that border one another. Note that this means that images of different parity that touch each other, at a critical curve, will be classified as one image. The points on the edges
or borders of each image are also found.  The inner border includes points
inside the image and the outer border is the points outside the image, but
bordering on it. The code goes through repeated grid
refinements until a termination criterion is met.  In each refinement, only the
grid points within the image or its outer border with the largest grid size
are refined. First a uniform grid refinement -- all cells within the image and outer border -- is 
done which ensures that each detected image has at least 50 points in it.

There are several choices for the termination criterion depending on the
application.  Common to all of them is that the outer border points must be at
least as well refined as the inner border points for each image.  This allows
the refined grid to expand out along narrow images whose boundaries were not well charactorized in the initial refinement steps.  If all that is required is an image with a certain
resolution the grid can be refined until all the points within the image have
that resolution.  If we are interested in the magnification ratios of images
it is desirable that the area of each image be calculated with equal
fractional accuracy.  In this case, the grid is refined until the largest cell
in the image border is a small fraction of the total area of that image, typically
$< 5.0\times 10^{-4}$.  This results in the grid refining to smaller scales
around smaller images.  Some images might be so small that they are
unobservable as in the central image of a singular mass distribution.  For
this reason, an additional minimum grid size requirement is imposed so that the
code does not refine indefinitely.  In other cases, such as microlensing, the
images are observationally indistinguishable and only the total area of all
images is important.  In this case, the same fractional criterion is 
applied to the images in total or the macro-image.

Each refinement of the grid requires a batch of evaluations of the lensing
quantities.  These are calculated in parallel as mentioned in section~\ref{sec:calc-defl}.  This makes the
timing scales almost linearly with the number of cores until the
neighbors-of-neighbors and image border search become the bottleneck.  For a simple 
analytic lens model the calls to the deflection solver can be very fast in which case the grid refinement is the bottleneck.  When there are many lens planes with many halos on each many cores ($\simgt 20$) can be used before this bottleneck is reached.

A way to reduce the grid refinement bottleneck is to brake the image plane up into regions, 
do the image finding and grid refinement in them separately.  We are so far able to do this 
for widely spaced patches of sky whose projections onto the source plane do not overlap.  
This mode of parallelization will be further developed in the future. 

\subsection{continuous surface brightness sources}

Realistic sources will not be circularly symmetric or constant surface brightness.  
For a varying surface brightness the refinement strategy must be modified.  The same telescoping and initial refinement is done with a circular source that is known to circumscribe the regions of the source that has significant flux.  The edge refinements are not done.  Instead a refinement scheme based on the surface brightness of the source is employed.  We have tested several schemes for this and found two to be useful.  The first is to simply require that the surface brightness in every cell not differ from the surface brightness in all its neighbors by more than a fixed threshold.  This makes smooth images but can sometimes over refine peaked surface brightness profiles.  Another strategy is to estimate the Gaussian curvature of the surface brightness at each cell by comparing it to its neighbors.  From the curvature an estimate of the error in the flux in that cell is made.  This error is then compared to the total flux of the image or all the images and refinement is stopped when this goes below a tolerance (typically $10^{-4}$).  This second method does a good jobs of measuring the total flux in an image, but occasionally when making images of realistic galaxies the outskirts of the galaxy, which contain a small fraction of the flux, can be less smooth than is desired.  We retain the two methods as options and continue to seek a more perfect solution.

\subsection{mapping critical curves and caustics}

It is often desired to calculate the caustic curves and their
images, the critical curves.  This is done by a modification of the process
used to finding images.  The regions of negative magnification are treated as
images.   Since points on either side of a critical curve have magnifications
of different sign the borders of all the negative magnification regions are the
critical curves.  Instead of refining the whole image region as for the images
only the borders are refined.  This is done until the required resolution is
reached.  The corresponding caustic curve is automatically found in the process since the critical curve is its image.  
The points in the curves are separated into separate curves by a neighbors-of-neighbor 
algorithm and then each curve is ordered to make a continuous curve. 
The critical curves are always closed curves or lead to the border of
the region being simulated.

\subsection{further comparison between the tree and FFT algorithms}

A final note on the relative scaling of FFT method versus the tree-code-AMR method employed within GLAMER, as noted before the bottleneck for a FFT code is generally putting the mass onto a uniform grid which scales as $N_{\rm grid} N_{\rm smooth}$ where $N_{\rm grid}$ is the number of grid points and $N_{\rm smooth}$ is a representative number of grid points within a smoothing length of a grid point, in projection.   Within strong lenses, if accuracy at high magnifications is required, a large amount of smoothing is often necessary and $N_{\rm grid}$ could be 10,000 or larger.  Thus this part of the code scale less well than sorting the particles into a tree ( $N_{\rm part} \log N_{\rm part}$).  Calculating the force in the tree code scales like $\sim N_{\rm grid}^* \log N_{\rm part}$ where $N_{\rm grid}^*$ is the number of grid points in the AMR grid which can be orders of magnitude smaller than the one for an FFT grid to reach the same resolution.  It is true that one can interpolate to smaller scales from the FFT uniform grid, but it is required that the initial grid spacing be smaller than the smallest smoothing length (zero in the case of stars).

Once the deflection angles have been found in the FFT approach the image finding is fairly trivial if it has been done to high enough resolution - simply find which source points lie within the source.  In this way the hard work can be done up front and the source can be changed at will later.  The difficulties comes in when many lenses configurations and/or projections are required, or extreme dynamical range or very small particles are required.

\begin{figure*}
\centering
\begin{minipage}{160mm}
\includegraphics[width=0.5\textwidth]{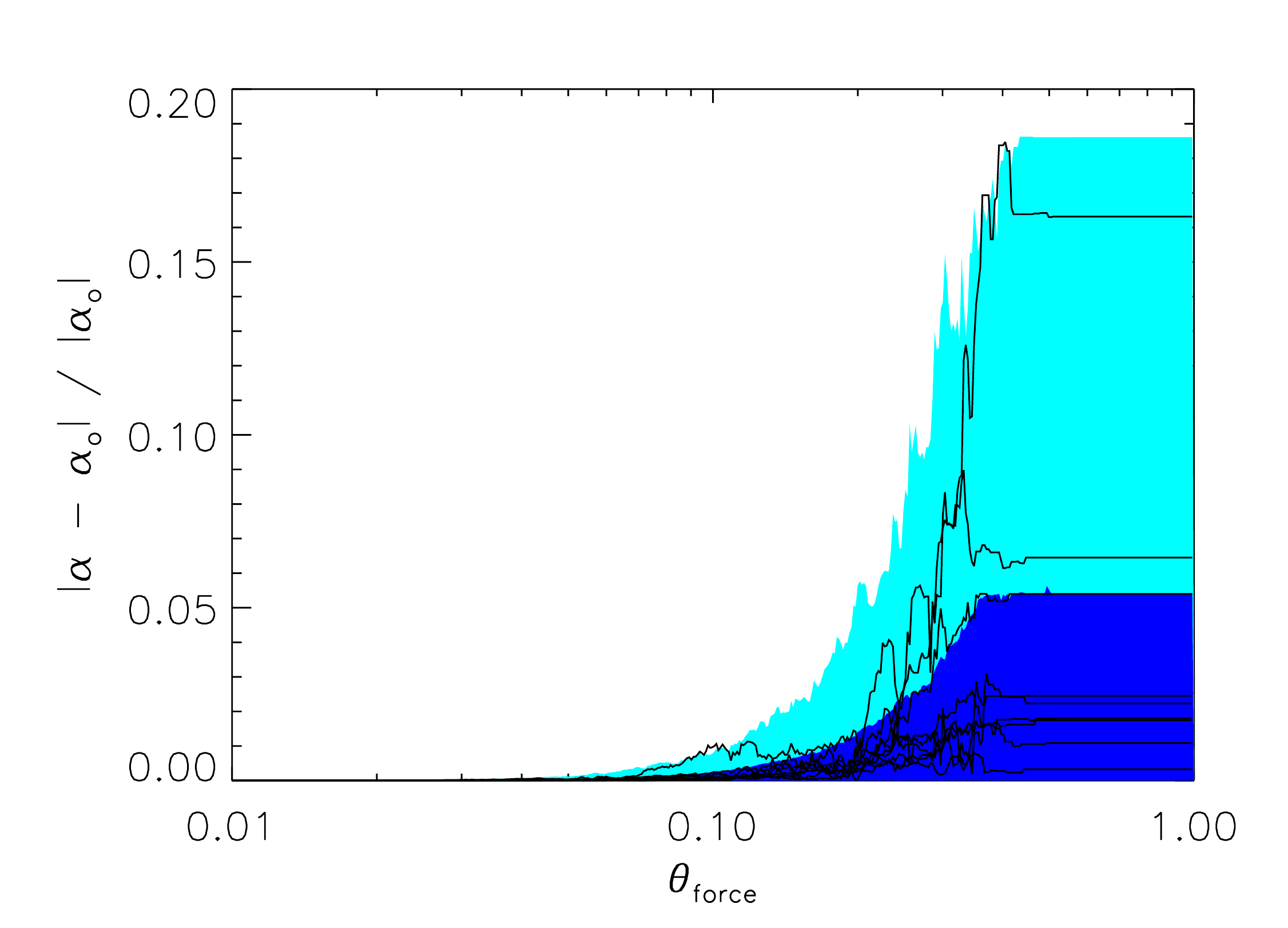}
\includegraphics[width=0.5\textwidth]{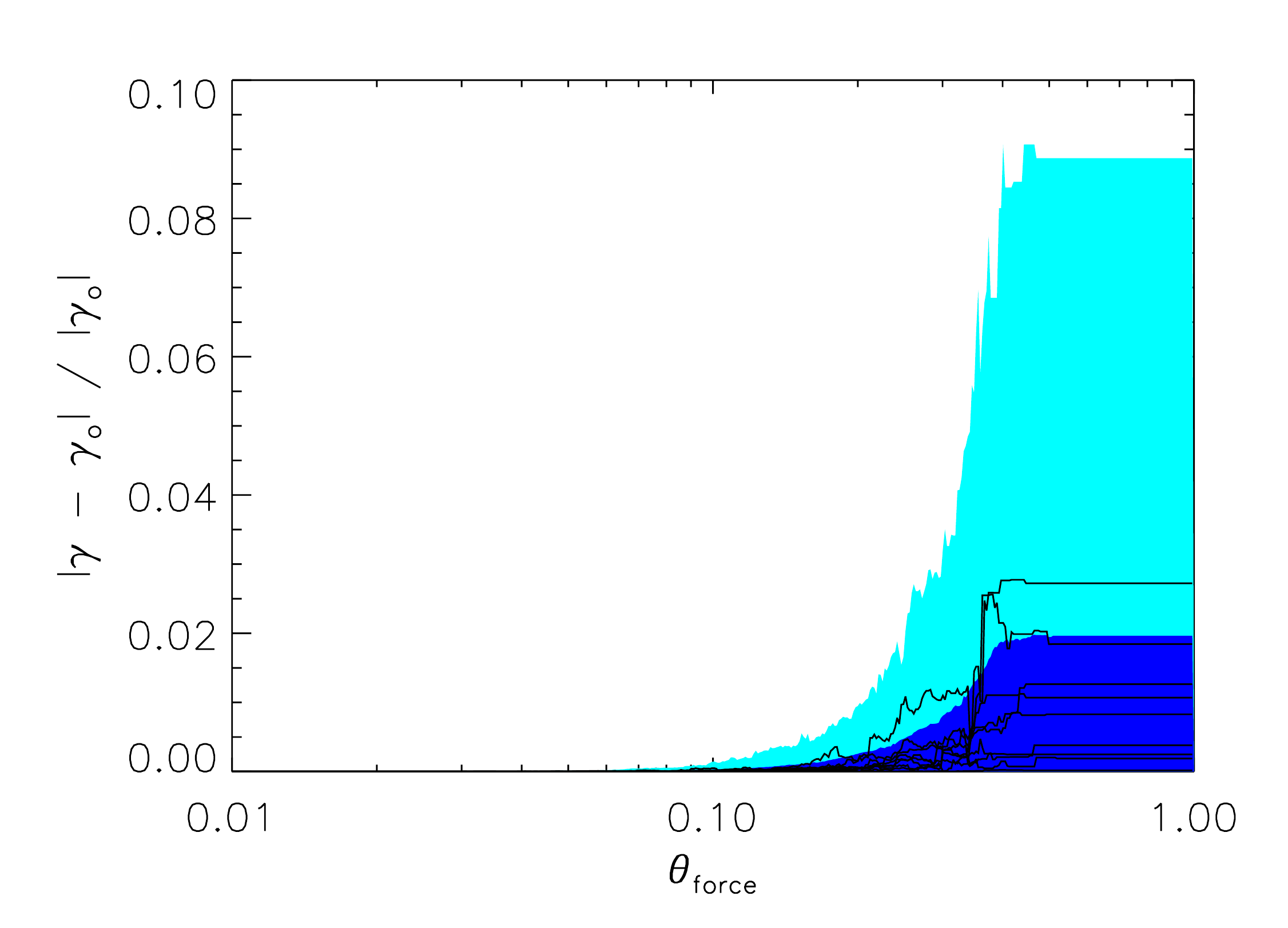}
\caption[]{ \label{fig:pm_theta_test}
Convergence test for tree deflection solver in the case of point
masses.  Within a circular region 10,000 point masses are placed
randomly and a ray is shot through the center of the circle.   The
parameter $\theta_{\rm force}$ (equation~\eqref{r_cut}) is increased from 0 to 1.  This is
repeated 1,000 times with different positions for the masses.  On the
left is the deflection angle ${\pmb \alpha}$
relative to the deflection angle for $\theta_{\rm force} = 0$, direct
summation case, the fractional error caused by the tree calculation.
On the right is the same for the shear, ${\pmb \gamma}$.  The blue
regions are where 90\% of the cases fall and the cyan regions are
where 99\% of the cases fall.  The curves are the first 10
realizations of the masses.
}
\end{minipage}
\end{figure*}

\begin{figure*}
\centering
\begin{minipage}{160mm}
\includegraphics[width=0.5\textwidth]{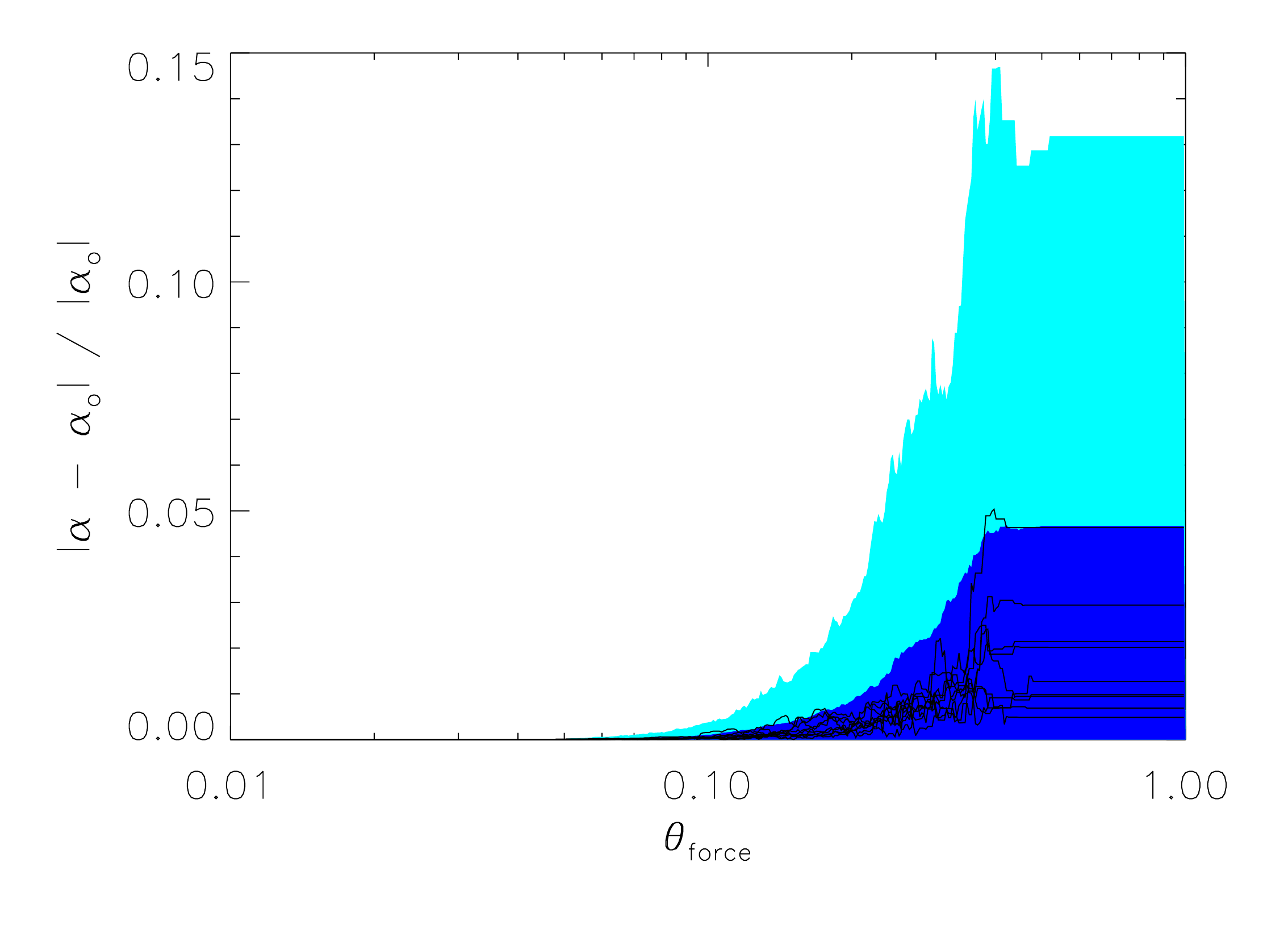}
\includegraphics[width=0.5\textwidth]{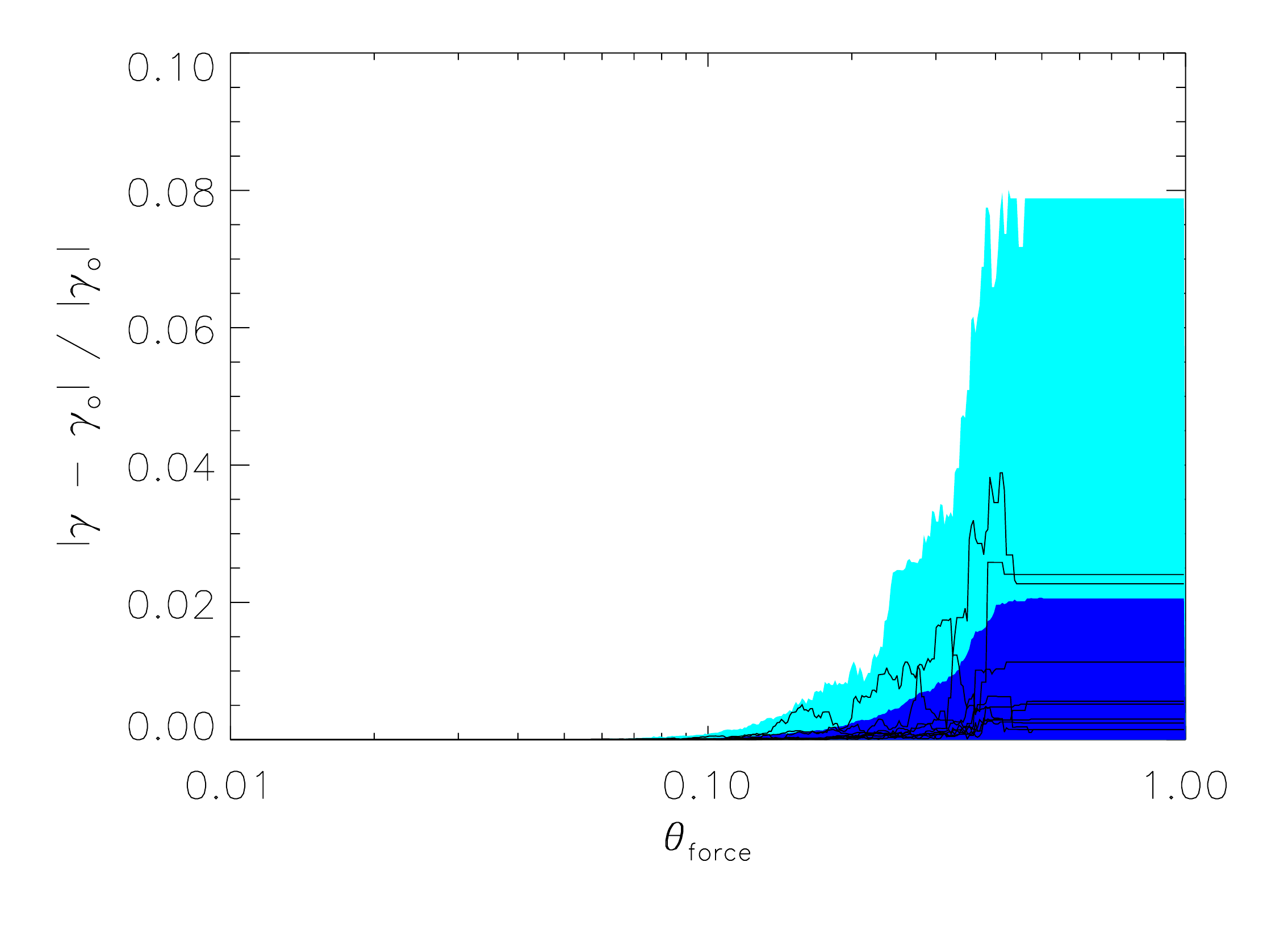}
\caption[]{ \label{fig:nfw_theta_test2}
Convergence test for tree deflection solver in the case of point
masses.   There are 1,000 point masses per realization.  There curves and
shaded regions are as in figure~\ref{fig:pm_theta_test}.}
\end{minipage}
\end{figure*}

\begin{figure*}
\centering
\begin{minipage}{160mm}
\includegraphics[width=0.5\textwidth]{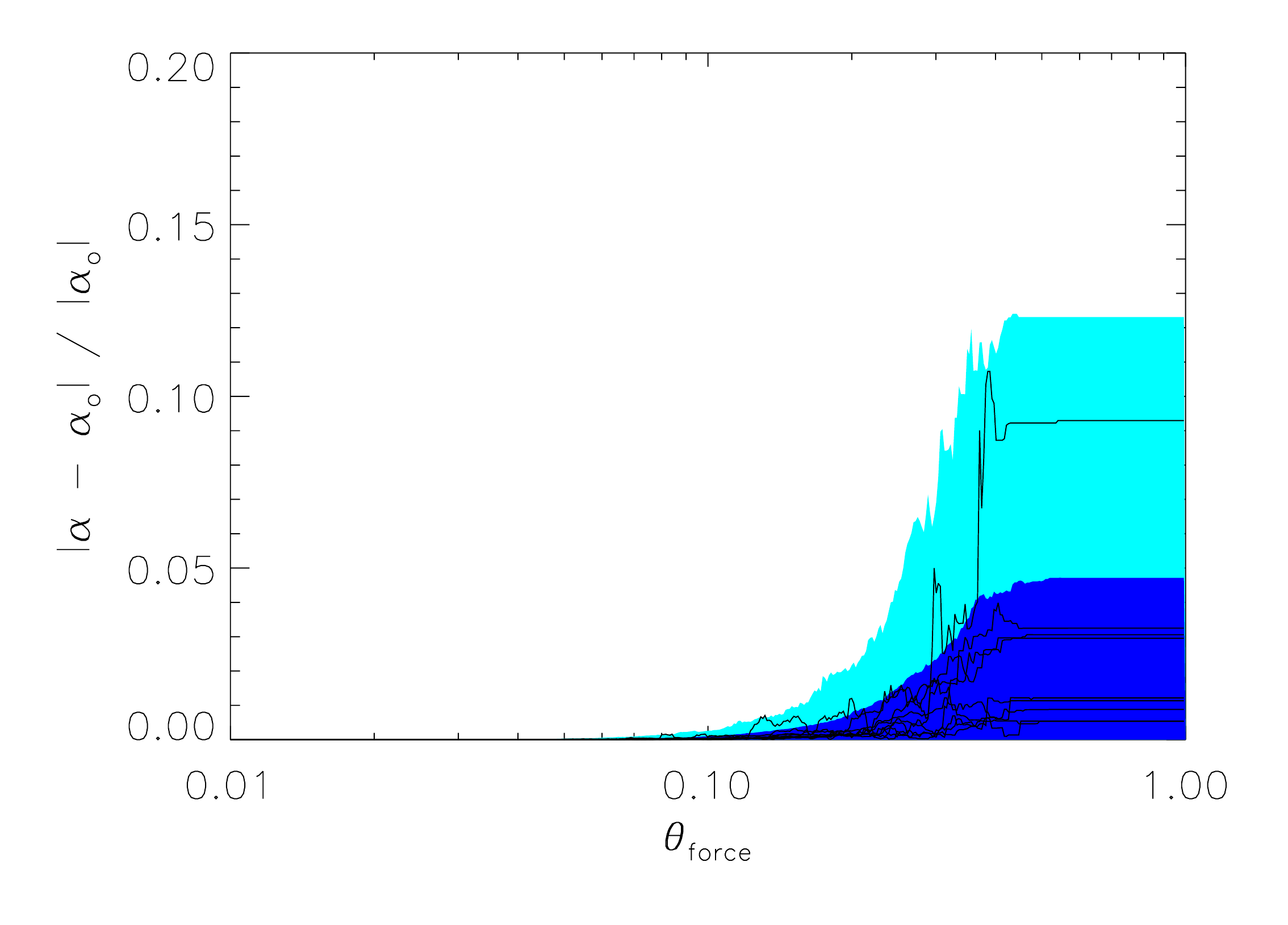}
\includegraphics[width=0.5\textwidth]{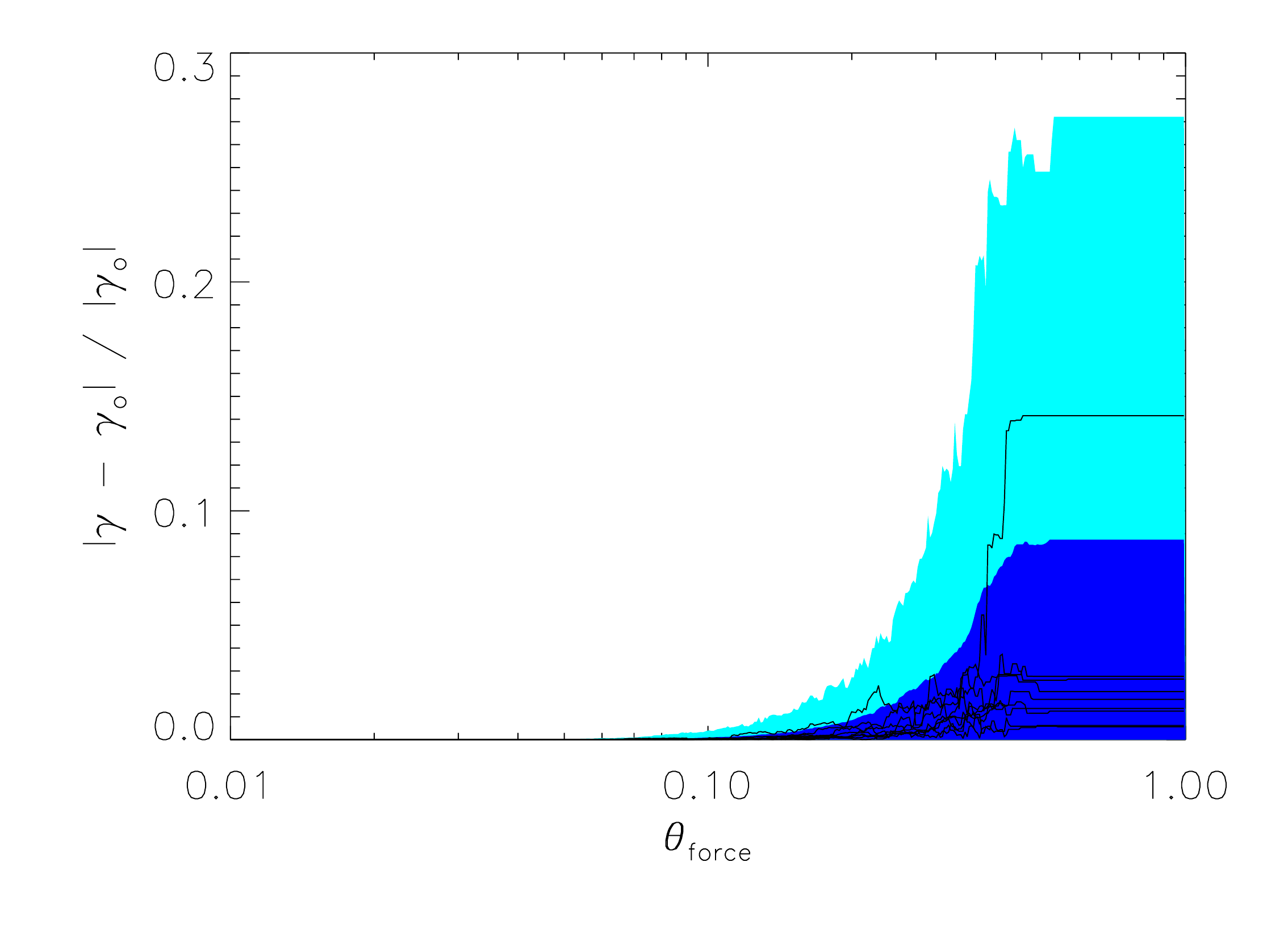}
\caption[]{ \label{fig:nfw_theta_test}
Convergence test for tree deflection solver in the case of halo
masses.  The halo properties are distributed as described in the
text.   There are 10,000 halos per realization.  There curves and
shaded regions are as in figure~\ref{fig:pm_theta_test}.}
\end{minipage}
\end{figure*}

\begin{figure*}
\centering
\begin{minipage}{160mm}
\includegraphics[width=0.5\textwidth]{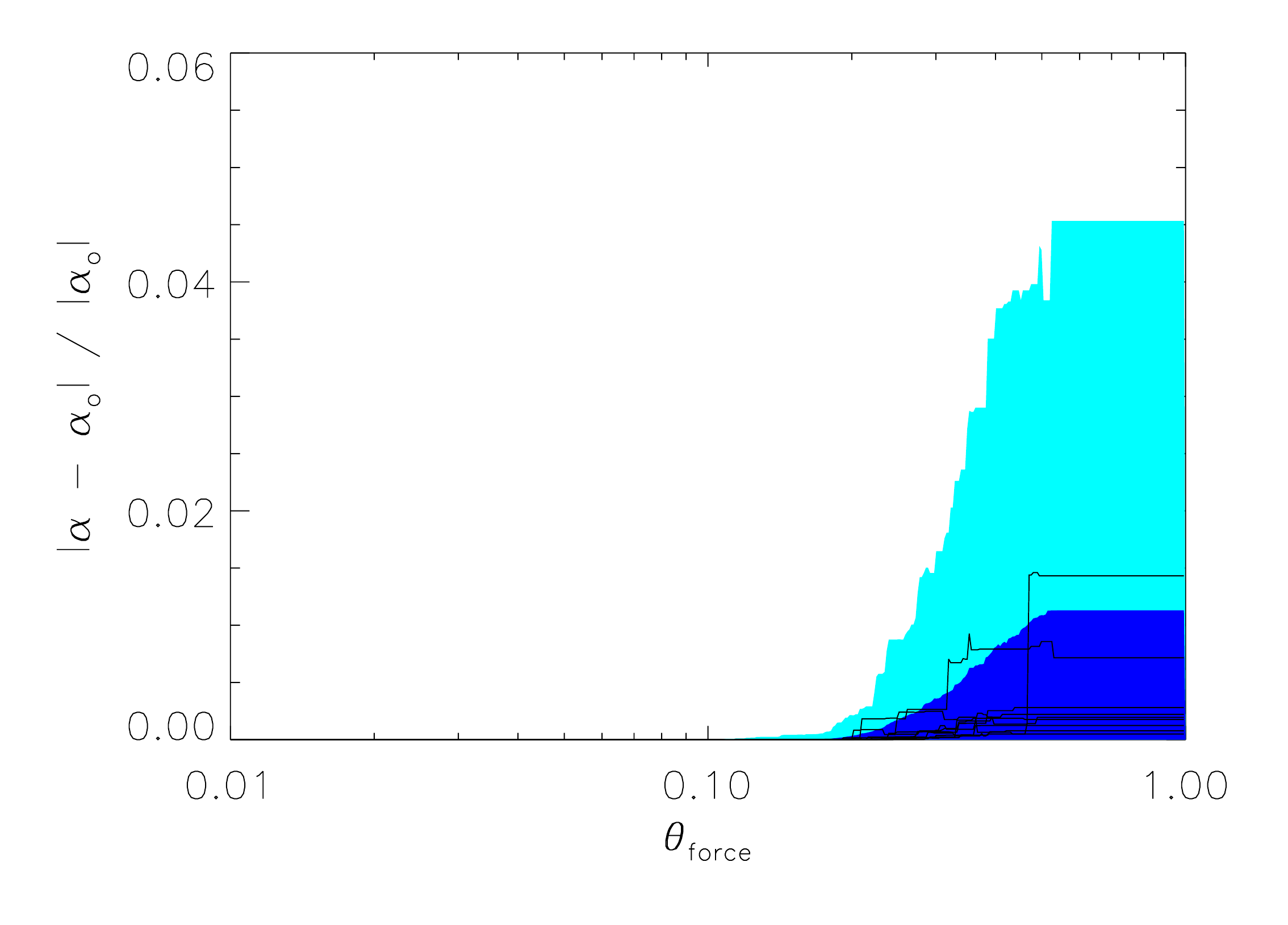}
\includegraphics[width=0.5\textwidth]{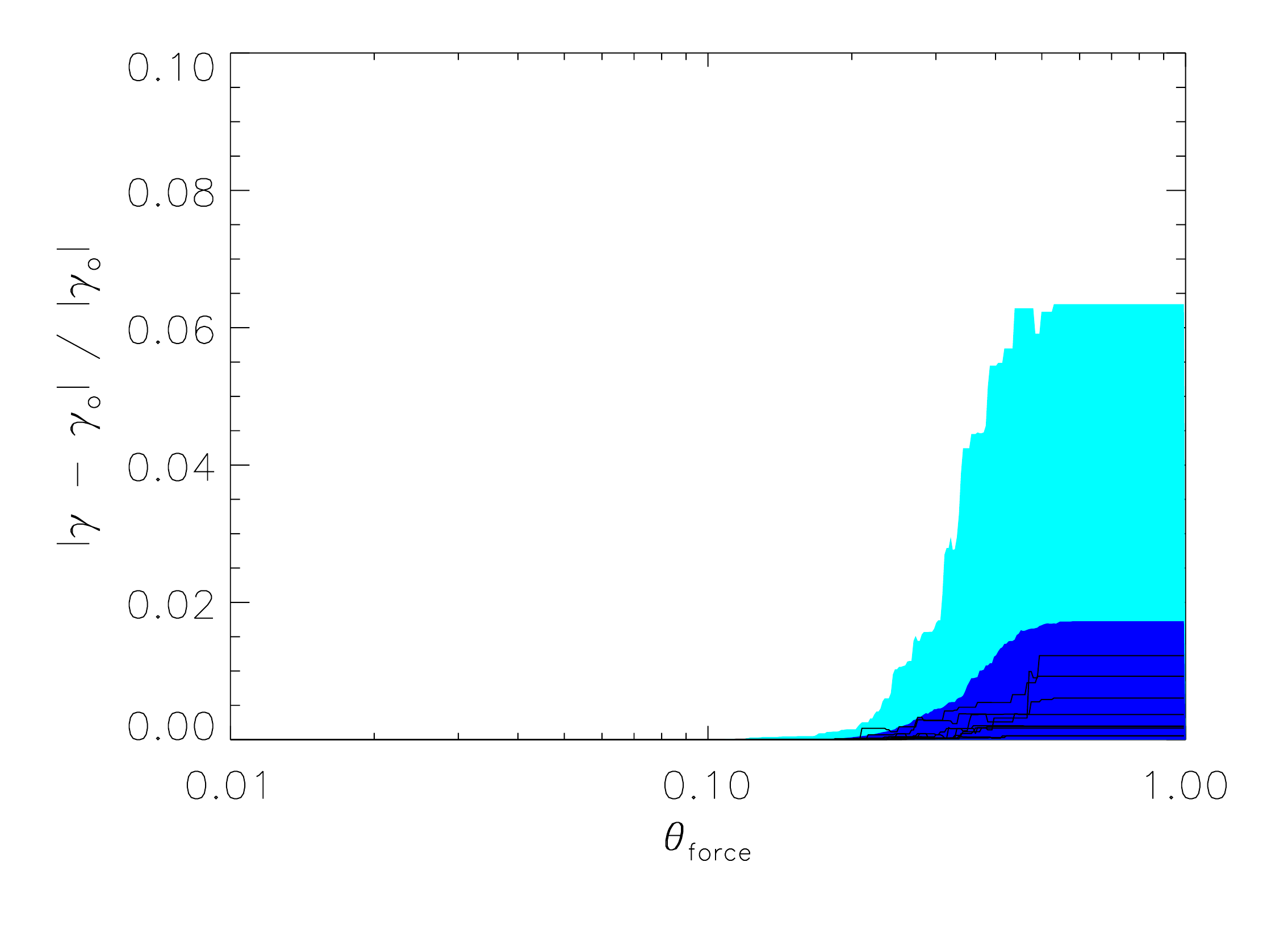}
\caption[]{ \label{fig:nfw_theta_lowd__test}
Convergence test for tree deflection solver in the case of halo
masses.  The halo properties are distributed as described in the
text.   There are 100 halos per realization.  There curves and
shaded regions are as in figure~\ref{fig:pm_theta_test}.}
\end{minipage}
\end{figure*}

\section{tests}
\label{sec:tests}

In this section, we briefly describe some of the tests we have
preformed on the  code.

\subsection{the tree deflection solver}
\label{sec:tree-defl-solv}
To test the tree deflections solver we have done a series of test some
of which will be presented here.  First we test the convergence of the
deflection solver's output with decreasing $\theta_{\rm force}$
(defined in \S~\ref{sec:calc-lens-quant}).   With $\theta_{\rm force}
=0$ the deflection solver adds the particles' contributions to the
lensing quantities one-by-one.  As $\theta_{\rm force}$ increases the
code uses the mass moments of larger and larger tree branches to calculate
these quantities and gets faster.

We start with random set of point masses in a circular region and
calculate the deflection and shear at the center with increasing
$\theta_{\rm force}$.   The background convergence, $\kappa$, is identically zero
in this case.  Figures~\ref{fig:pm_theta_test} and
\ref{fig:nfw_theta_test2} show how these
quantities converge to the $\theta_{\rm force}=0$ values as
$\theta_{\rm force}$ decreases.  The masses are given unit mass, the
radius of the circle is set to unity so that the effective optical
depths (the density of stars in units of their Einstien radius, 
$\Sigma_{\rm star} \pi r^2_ e / m_{\rm star} $) are $10^4$ and $10^3$ 
which are very high compared to those usually 
encountered in observations, but are extreme
cases that challenge the deflection solver.

We repeat the convergence test for our halo models.  Here we will show
the NFW halo.  Again a unit circle is populated with masses.  The sizes
of the halos are taken to be uniformly distributed up to 0.3.  The
masses are taken from a distribution $\frac{dn}{dm} \propto m^{-2}$
with a maximum mass of unity and a minimum of $10^{-3}$.  They are all 
given a concentration of 3.  The algorithm should identify all halos
that intersect the ray and calculate their ${\pmb \alpha}$, $\kappa$ and ${\pmb\gamma}$
by direct summation.   Since only halos that intersect the ray
contribute to $\kappa$ there should be no error for any choice of
$\theta_{\rm force}$ compared to the  $\theta_{\rm force}=0$.  We
verify that this is true.  The errors in ${\pmb \alpha}$ and
${\pmb\gamma}$ are shown in figure~\ref{fig:nfw_theta_test} for 10,000
halos within the circle.  We also repeated the exercise with 100 halos
per realization with smaller sizes (uniform up to 0.2) shown in 
figure~\ref{fig:nfw_theta_lowd__test}.

The convergence of ${\pmb \gamma}$ with $\theta_{\rm
  force}$ is generally better than for ${\pmb \alpha}$ because these
quantities fall off faster with distance between the ray and the
center of mass.  For this reason it is the accuracy of ${\pmb \alpha}$
that determines how large a $\theta_{\rm force}$ is acceptable.
  We conclude from these test and others that a value of $\theta_{\rm
  force} = 0.1$ is sufficiently small for most applications.

\subsection{singular isothermal sphere (SIS) tests}
\label{sec:sing-isoth-sphere}

\begin{figure}
\centering
\includegraphics[width=0.49\textwidth]{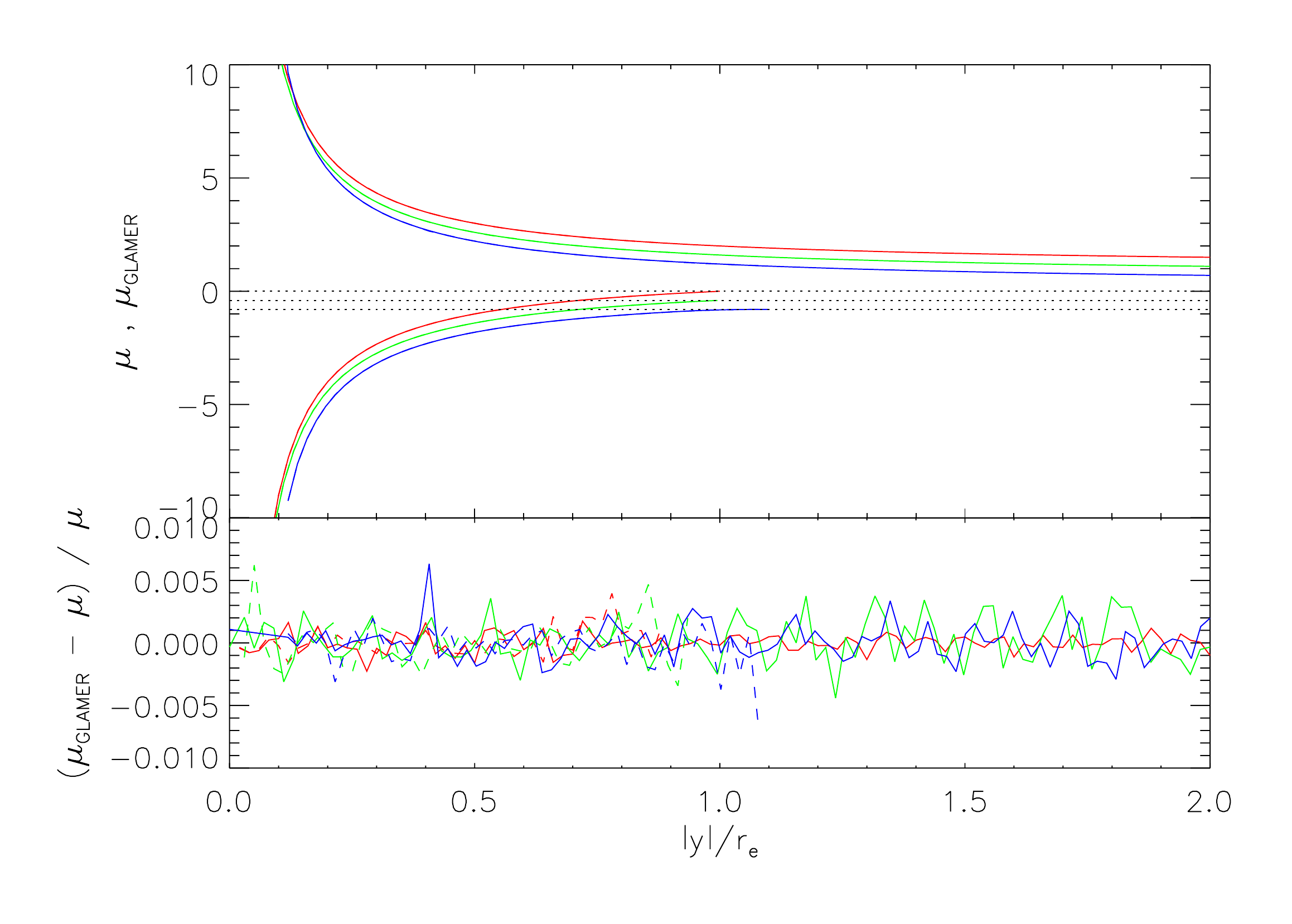}
\caption[]{ \label{fig:sistest}
A comparison of the analytically calculated image magnifications for a
Singular Isothermal Sphere (SIS) with those calculated with GLAMER.
Above are the magnifications for three different source sizes - 0.1 (blue),
0.01 (green) and 0.0001 (red) Einstein radii.  The results for each source size are
offset vertically so that they can be differentiated.   The dotted line marks
zero magnification in each case.  The negative magnification curves
are for the negative parity image that is closest to the center of the
lens.  This image does not appear when no part of the source is within
one Einstein radius, $r_e$, of the center of the lens.  Both the
analytic result (solid curves) and the GLAMER result (dashed curves)
are shown in the top panel, but they are hard to differentiate.  The
fractional difference between them is shown on the bottom with the
same color scheme.
}
\end{figure}
\begin{figure}
\centering
\includegraphics[width=0.46\textwidth]{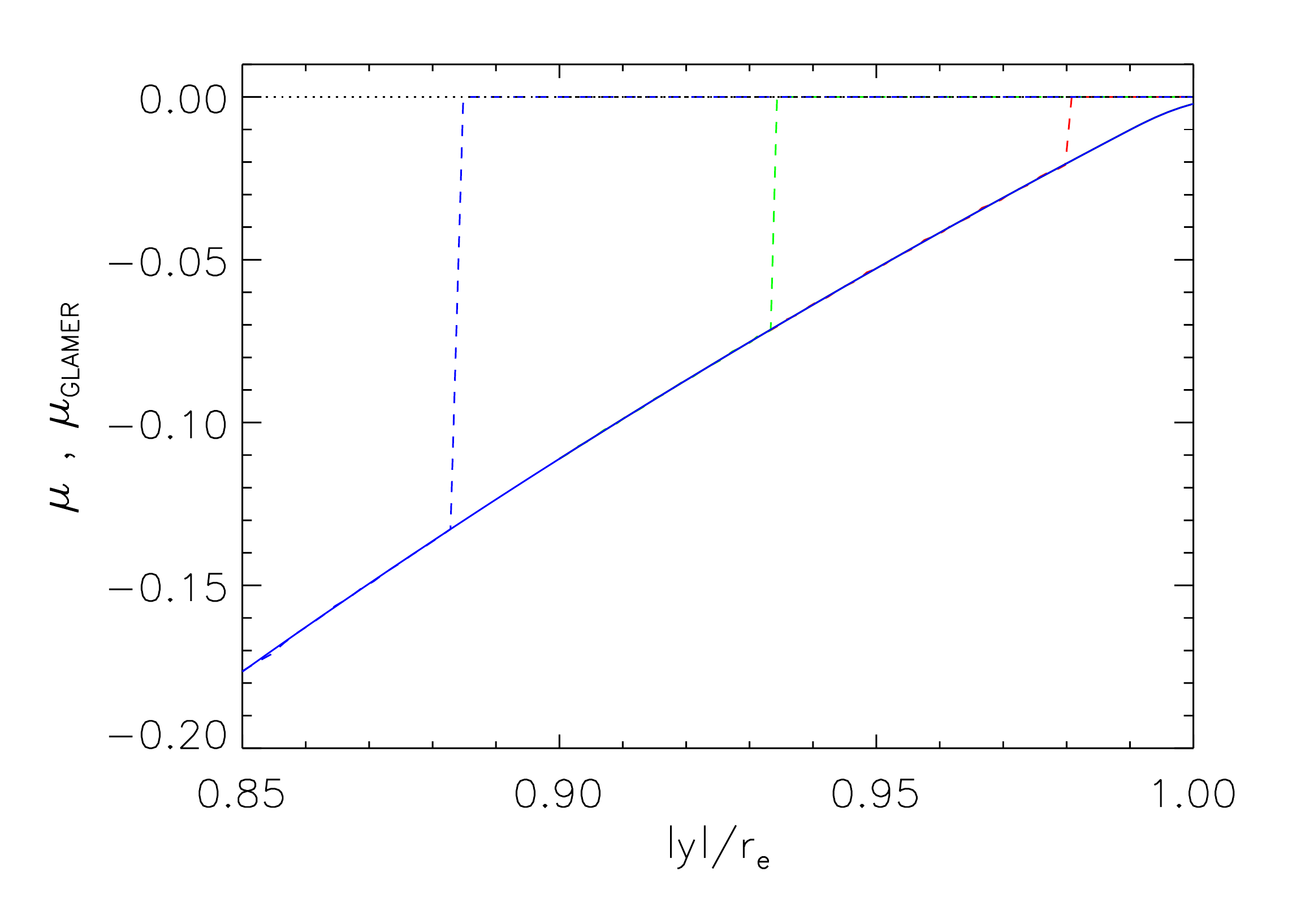}
\caption[]{ \label{fig:mumintest}
The magnification of the negative parity image in a SIS model
according to analytic calculation (solid curves) and the GLAMER code
(dashed curves) as a function of source position.  The source radius 
is $0.01 r_e$.  The GLAMER code looses the image and its magnification
goes to zero before the correct solution does.  The values of
$\mu_{\rm min}$ from left to right are 0.16 (blue), 0.09 (green) and
0.01 (red).  A $\mu_{\rm min}$ of 0.09 should catch all images with a
magnification above 0.07 for example.
}
\end{figure}

Lensing by a singular isothermal sphere is particularly easy to work
with analytically. 
Appendix~\ref{app:sing-isoth-sphere-1} summarizes some analytic
properties of a SIS lens and how to calculate the magnification of a
finite source with simple numerical integrations.   These solutions
can be compared to the results of GLAMER to test its grid refinement.

We set up an SIS lens with GLAMER for a lens at $z=0.34$, a source
at $z=3.62$ and $\sigma = 300~\kms$ although these parameters 
should not have any effect on the results since all quantities are a 
function of the $y/r_e$ where $r_e$ is the Einstein radius 
(equation~\eqref{eq:einstein_radius}).  With an initial 64-by-64 grid over an
area of 20 $r_e$ squared GLAMER finds the images and their
magnifications.  The center of the source, $y_c$ is moved from zero to 
$2 r_e$ for three source sizes.   The grid is erased and recreated for each source position.  Otherwise hysteresis in the grid refinement can help in finding small images.  The grid convergence requirement is set for each image individually.

The resulting magnifications are
compared with the analytic results in figure~\ref{fig:sistest}.  The
errors are generally below 0.5\%  and never greater than 0.7\%.   The
errors show no significant dependence on source size, image
magnification or image shape.   The smallest source tested here is
$10^5$ times smaller than the initial grid.

When the source passes outside of the circle $|y| < r_e$, the
magnification of the negative image goes to zero.  If the image is too
small relative to the source size GLAMER will miss the image because of the
finite resolution of the initial grid,  as explained in
section~\ref{sec:find-mapp-imag}.  Figure~\ref{fig:mumintest} shows
the region of the magnification curves where the negative image is
disappearing for a source size of $0.01 r_e$.  Decreasing the value of
the parameter $\mu_{\rm min}$ makes smaller images detectable at the
expense of shooting and sorting more rays.   In
most realistic cases there is at least one image with a magnification
greater than or equal to one \footnote{One notable exception is when
  simulating a single macro image of a larger strong lens separately
  as might be done for investigating microlensing.   In this case, the
  total magnification could be significantly less than one and special
care needs to be taken.   In cosmological simulations it is also
possible to get total magnifications below one for lines of sight that
pass through under dense regions.  In realistic cases, these are only
weakly demagnified and the same argument holds that the total
magnification (all images) of a source is $\simgt$ 1.} so $\mu_{\rm min}$ can also be
seen as controlling the smallest magnification ratio that is expected
to be found.  The  $\mu_{\rm min}$ is adjusted for the particular
application, but has a default value of 0.09.

\subsection{image finding and grid refinement}
\label{sec:imagefindingtests}

\begin{figure*}
\centering
\includegraphics[width=1.0\textwidth]{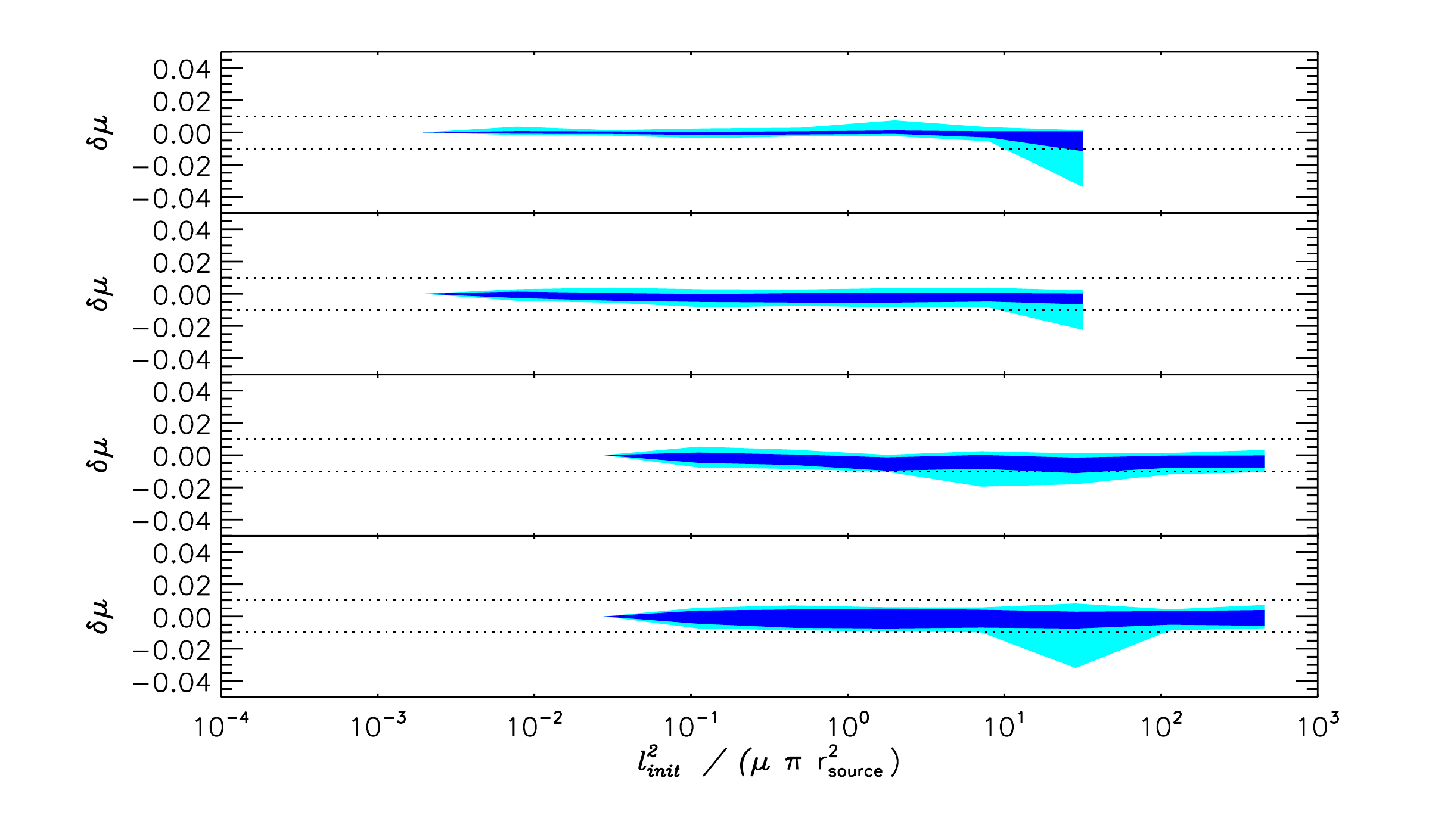}
\caption[]{ \label{fig:microtest}
A test for how well the adaptive mesh method tracks images.  A uniform distribution of 
stars with a background shear.  The initial grid resolution is denoted $l_{\rm init}$ 
and $\bar{\mu}$ is what the magnification would be if the mass in the stars where spread 
uniformly.  The error in the magnification caused by the adaptive mesh calculation
is defined as $\delta\mu = \left\{ \mu(l_{\rm init}) - \mu(l_{\rm
    init} = l_{\rm min}) \right\} /\bar{\mu}$ where $l_{\rm min}$ is
the smallest initial grid size used in the particular run.  The top
two panels are for $\bar{\kappa} = 0.45$ and $\bar{\pmb\gamma} = \{0,0\}$ and
-- from top to bottom -- source sizes $r_{\rm source} = 0.001~\pc$ and
0.01~pc.  The Einstein radius of the stars in these cases is
0.175~pc.  The bottom
two panels are for $\bar{\kappa} = 0.54$ and $\bar{\pmb\gamma} = \{0,0.4\}$ and
with the same source sizes.  The dark blue region contains 90\% of the
cases at each initial grid size and the light light blue regions shows the envelope of all the realizations.  There are typically hundreds of micro-images in these 
simulations.  For each panel there are 100 realizations of the stellar distribution 
except in the second from top where there are 120.
}
\end{figure*}

\begin{figure*}
\centering
\includegraphics[width=1.0\textwidth]{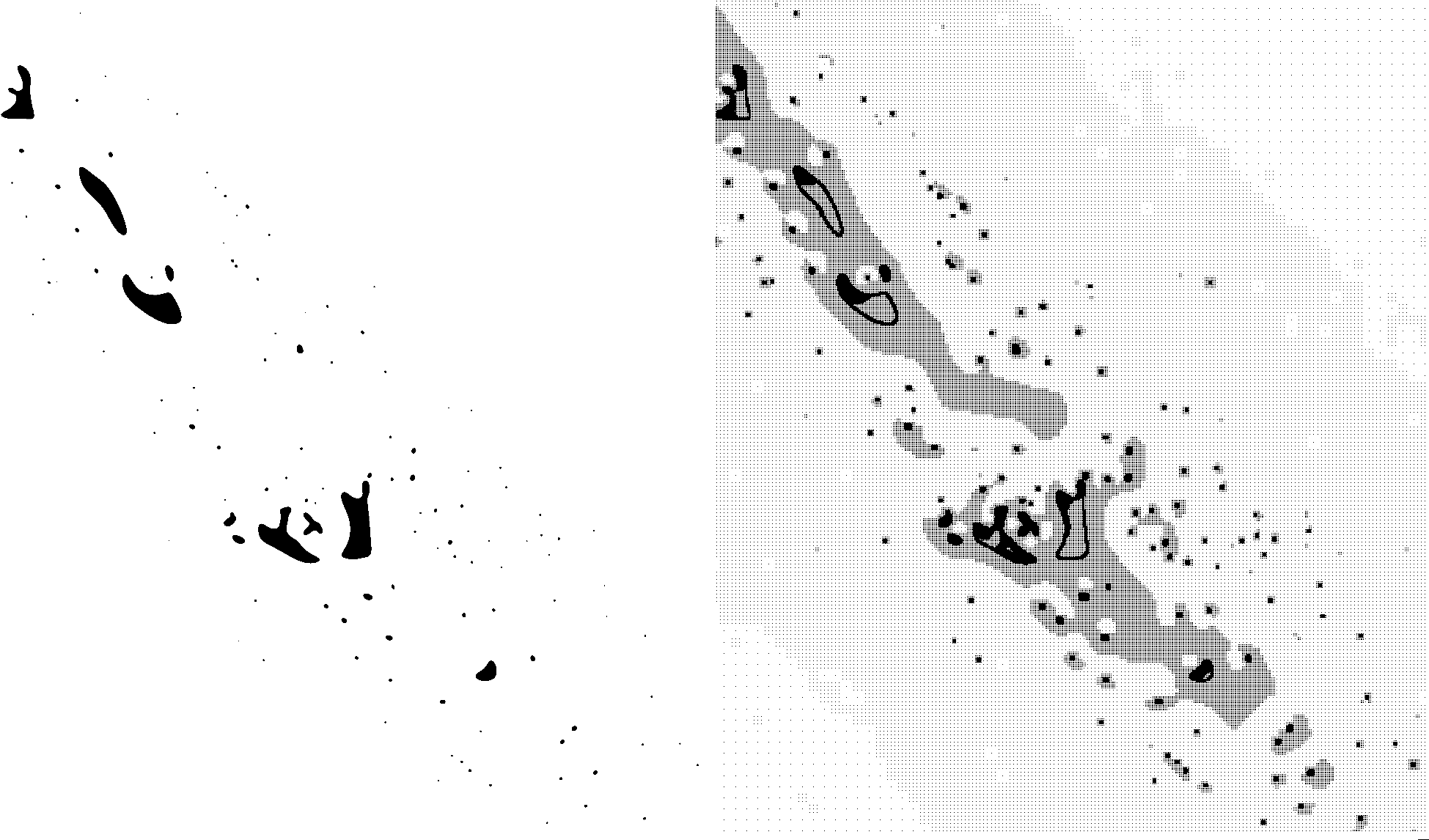}
\caption[]{ \label{fig:image_and_grid}
On the left is a detail of the images produced for a circular  source lensed by stars with an average convergence in stars of $\bar{\kappa} = 0.54$ and a background shear of $\bar{\pmb\gamma} = \{0,0.4\}$.  On the right is the grid or ray pattern created during the refinement process to these images.  There is a dot at each ray, although the pixelization of the image might make this hard to see. The grid was initialized in a field two orders of magnitude larger than this one with 16 x 16 pixels. The images extend well beyond the pictured region. It can be seen that for some images just the edges are refined to the highest visible resolution.
}
\end{figure*} 

\begin{figure*}
\centering
\includegraphics[width=1.0\textwidth]{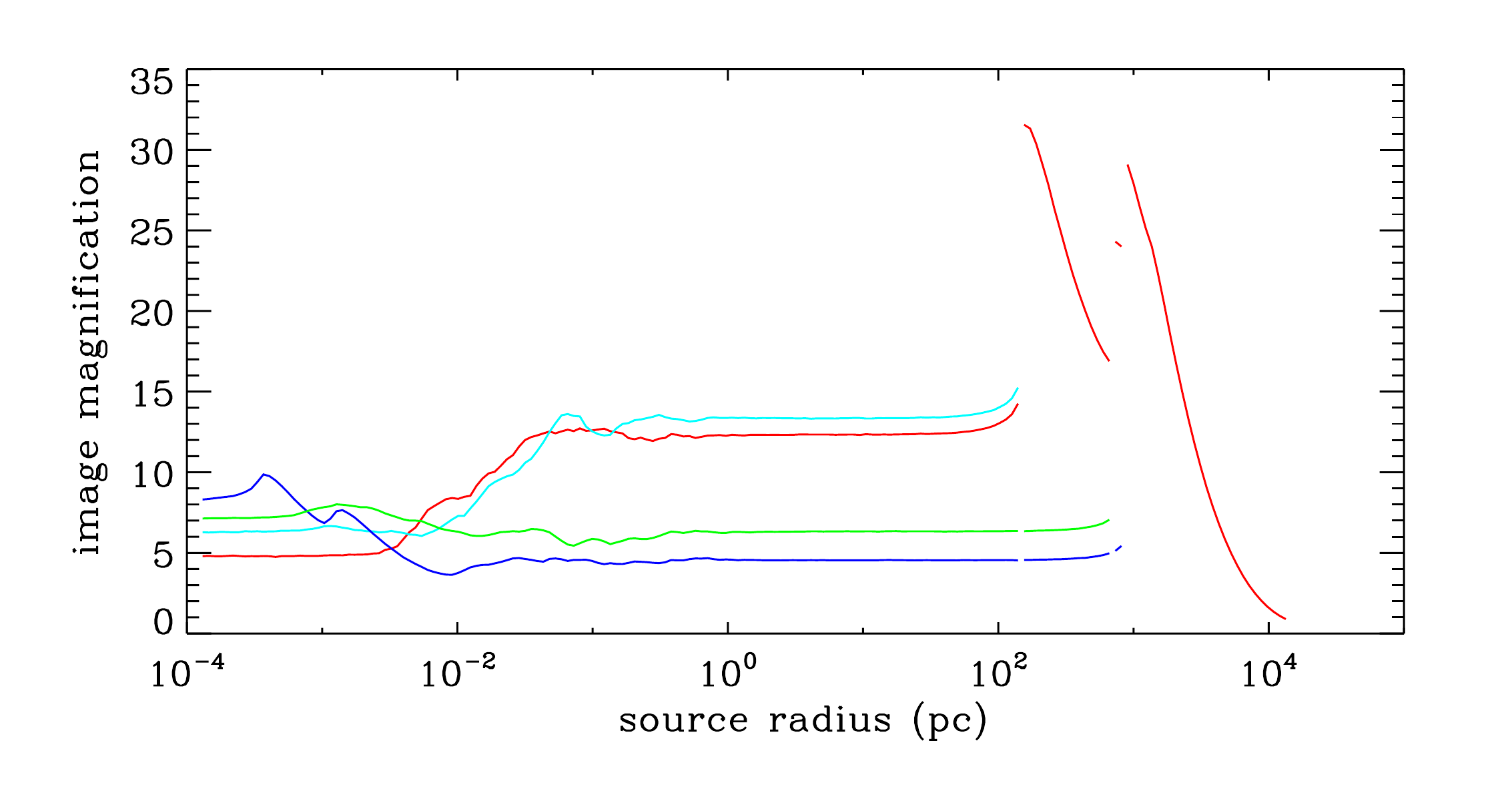}
\caption[]{ \label{fig:fullrange}
The magnification as a function of source size for a source at z=3.62  and a Singular Isothermal Ellipsoid (SIE) lens with stars in it at z=0.3.  The SIE has a velocity dispersion of $\sigma =300$~km/s.  At the positions of the images all the mass is in stars.  For the largest source sizes the image is larger than the lens and has a magnification of about one.  As the source gets smaller it forms an Einstein ring with high magnification.  The ring splits into two (for a very short range), then three and then four as the source gets smaller.  When the source gets below 1 parsec in radius the four macro-images start to break up into micro-images and the magnifications start to fluctuate because of microlensing.
}
\end{figure*} 

Quasar microlensing is the most challenging case for any ray-shooting
code.  In this case, a high redshift quasars that is being strongly lensed by an intervening galaxy is also being affected by individual stars in that lens.   The dynamic range is very large (tens of kpc down to $\sim 10^{-5}$~pc) and many (hundreds) of irregular shaped micro-images can be present.  Because the deflection caused by a (Newtonian) point-mass diverges, each star is capable of producing an image no matter how far away it is from the direct line-of-sight to the source.  But images around stars that are very far from the line-of-sight will be very small.  It is not practical, or desirable, to find and map every image because some will be insignificant to the total magnification of the macro-image. The challenge here lies in finding the micro-images that contribute while shooting as few rays as possible.  It has already been demonstrated that the code can calculate the magnification of an individual image accurately once it is found.

To test this aspect of the code we compare the magnifications calculated 
with adaptive mesh refinement to those calculated with a high
resolution initial grid that is guaranteed to find images smaller than
a certain size even before any refinement has been done.  We use what we call a uniform lens for this which
initially has a uniform convergence and shear, $\bar{\kappa}$ and
$\bar{\pmb \gamma}$.  Stars are then implanted randomly in a circular
region and the corresponding mass is subtracted from the initial mass
sheet.  For these trials we use 10,000 stars.  Each trial starts with
a very coarse initial grid of 16 by 16 rays and increases the initial
grid resolution until a point where small images would be identifiable
on the initial grid without any refinement.  The initial grid
resolution is denoted $l_{\rm init}$.  Getting a consistent result
independent of the initial grid resolution indicates that the adaptive
refinement is doing its job well and gives an indication of the error
in the magnification.  This is repeated 100 times with
different star positions, but the same $\bar{\kappa}$ and $\bar{\pmb
  \gamma}$.  Then it was repeated for different $\bar{\kappa}$,
$\bar{\pmb \gamma}$ and source sizes.  Figure~\ref{fig:microtest}
shows some of these tests. 

The code is usually able to measure the total magnification of the
macro-image to better than 1\%.  There are some outliers where
the error is as big as 4\%, but less than 1\% of the time.   This
seems sufficiently accurate considering that typical observational
uncertainties are $\sim 10\%$.  More accuracy is attainable by
decreasing $\mu_{\rm min}$ at the expense of increasing the
computational time.

Figure~\ref{fig:image_and_grid}, shows a portion of the macro-image produced in a simulation along with the 
grid pattern that the code chose to find its area.  It can be seen how the code locates the images and then tries 
to refine around the edges of the images in the case of a uniform surface brightness circular source.  This is accomplished even in the case of many very irregularly shaped images.   

In figure~\ref{fig:fullrange} the code's ability to cover a very large
range of angular scales is demonstrated.  Here the image
magnifications are plotted for a source as a function of its radius
from super-galactic scales down to $10^{-4}$ parsecs scales.  A galaxy
sized SIE (single isothermal ellipsoid) lens ($\sigma = 300$~km/s) is
included and 10,000 stars are implanted in regions surrounding each
macro-image.  The locations for implanting the stars are found by
initially finding the image positions with the smooth host model
without stars.  All the mass is in stars at the image positions.
Distinct ranges in source size can be seen in the plot where the host
lens is important, where stars are important to the magnifications and
where the source is even smaller than any structure in the
microlensing magnification map.  The magnification curves have been calculated both from an adapted grid initialized with a size of 5~arcsec and with four separate grids initialized with a size of $\sim 10$~pc on the lens plane centered on each of the macro-images.  The same calculation has been preformed with different values for $\mu_{\rm min}$ and numbers of points in the initial grid to confirm consistency of the magnification curves.  

Countless other checks have been made during the development of the
code.  For example, all the  analytic halo profiles have been checked
and the interpolation has been compared to directly ray-shooting the points and
found to agree to a part in $10^4$ for the deflection.   We have also constructed approximations to analytic lens profiles out of particles and tested that their lensing properties agree with analytic results.

\section{discussion}
\label{sec:discussion}

We have succeeded in developing an adaptive lensing code that is flexible enough with its representation of sources and lenses that it can be used for many different applications.  This code is capable of simulating everything from QSO microlensing to weak shear on multi-degree scales while reconstructing individual images.  
The lensing mass can be represented by point masses (stars), smoothed simulation particles, analytic halo models, pixelized mass maps or any combination of these and easily switch between them.  Likewise the source can be represented by a uniform surface brightness circle, an analytic surface brightness model or a pixelized image.
 
We continue to develop GLAMER to make it a more useful tool.  In a companion paper \citep{PM&G2013} we describe how the code has been extended to allow for lensing through three dimensional mass distributions so that whole light cones can be simulated.  We continue to add to the number of source models available.  More sophisticated methods of representing galaxies and QSO emission regions are being implemented.  More sophisticated halo models are being implemented with a variety of profiles and asymmetries.  Finally, the machinery used to do simulations is being adapted to do lens fitting for weak and strong lensing.

\vspace{0.3cm} 
\leftline{\bf Acknowledgments} 
We would like to thank our colleagues in Bologna -- F. Bellagamba, C. Giocoli, D. Leier and N. Tessore -- for helpful contributions and for their further development of the code that will appear in future publications.  This research is part of the project GLENCO, funded under the Seventh Framework Programme, Ideas, Grant Agreement n. 259349.

 \bibliographystyle{/Users/bmetcalf/Work/TeX/apj/apj}
 \bibliography{/Users/bmetcalf/Work/mybib}

\begin{thebibliography}{23}
\expandafter\ifx\csname natexlab\endcsname\relax\def\natexlab#1{#1}\fi

\bibitem[{{Amara} {et~al.}(2006){Amara}, {Metcalf}, {Cox}, \&
  {Ostriker}}]{AMCO04}
{Amara}, A., {Metcalf}, R.~B., {Cox}, T.~J., \& {Ostriker}, J.~P. 2006, \mnras,
  367, 1367

\bibitem[{{Angulo} {et~al.}(2013){Angulo}, {Chen}, {Hilbert}, \&
  {Abel}}]{2013arXiv1309.1161A}
{Angulo}, R.~E., {Chen}, R., {Hilbert}, S., \& {Abel}, T. 2013, ArXiv e-prints

\bibitem[{{Aubert} {et~al.}(2007){Aubert}, {Amara}, \&
  {Metcalf}}]{2007MNRAS.376..113A}
{Aubert}, D., {Amara}, A., \& {Metcalf}, R.~B. 2007, \mnras, 376, 113

\bibitem[{{Barnes} \& {Hut}(1989)}]{1989ApJS...70..389B}
{Barnes}, J.~E. \& {Hut}, P. 1989, \apjs, 70, 389

\bibitem[{{Fluke} {et~al.}(1999){Fluke}, {Webster}, \&
  {Mortlock}}]{1999MNRAS.306..567F}
{Fluke}, C.~J., {Webster}, R.~L., \& {Mortlock}, D.~J. 1999, \mnras, 306, 567

\bibitem[{{Hamana} \& {Mellier}(2001)}]{2001MNRAS.327..169H}
{Hamana}, T. \& {Mellier}, Y. 2001, \mnras, 327, 169

\bibitem[{{Hilbert} {et~al.}(2009){Hilbert}, {Hartlap}, {White}, \&
  {Schneider}}]{2009A&A...499...31H}
{Hilbert}, S., {Hartlap}, J., {White}, S.~D.~M., \& {Schneider}, P. 2009, \aap,
  499, 31

\bibitem[{{Hilbert} {et~al.}(2007{\natexlab{a}}){Hilbert}, {White}, {Hartlap},
  \& {Schneider}}]{hilbert2007}
{Hilbert}, S., {White}, S.~D.~M., {Hartlap}, J., \& {Schneider}, P.
  2007{\natexlab{a}}, ArXiv Astrophysics e-prints, astro-ph/0703803

\bibitem[{{Hilbert} {et~al.}(2007{\natexlab{b}}){Hilbert}, {White}, {Hartlap},
  \& {Schneider}}]{2007MNRAS.382..121H}
---. 2007{\natexlab{b}}, \mnras, 382, 121

\bibitem[{{Jain} {et~al.}(2000){Jain}, {Seljak}, \&
  {White}}]{2000ApJ...530..547J}
{Jain}, B., {Seljak}, U., \& {White}, S. 2000, \apj, 530, 547

\bibitem[{{Killedar} {et~al.}(2012){Killedar}, {Lasky}, {Lewis}, \&
  {Fluke}}]{2012MNRAS.420..155K}
{Killedar}, M., {Lasky}, P.~D., {Lewis}, G.~F., \& {Fluke}, C.~J. 2012, \mnras,
  420, 155

\bibitem[{{Li} {et~al.}(2006){Li}, {Mao}, {Jing}, {Kang}, \&
  {Bartelmann}}]{2006ApJ...652...43L}
{Li}, G.-L., {Mao}, S., {Jing}, Y.~P., {Kang}, X., \& {Bartelmann}, M. 2006,
  \apj, 652, 43

\bibitem[{{Mediavilla} {et~al.}(2006){Mediavilla}, {Mu{\~n}oz}, {Lopez},
  {Mediavilla}, {Abajas}, {Gonzalez-Morcillo}, \&
  {Gil-Merino}}]{2006ApJ...653..942M}
{Mediavilla}, E., {Mu{\~n}oz}, J.~A., {Lopez}, P., {Mediavilla}, T., {Abajas},
  C., {Gonzalez-Morcillo}, C., \& {Gil-Merino}, R. 2006, \apj, 653, 942

\bibitem[{{Meneghetti} {et~al.}(2008){Meneghetti}, {Melchior}, {Grazian}, {De
  Lucia}, {Dolag}, {Bartelmann}, {Heymans}, {Moscardini}, \&
  {Radovich}}]{2008A&A...482..403M}
{Meneghetti}, M., {Melchior}, P., {Grazian}, A., {De Lucia}, G., {Dolag}, K.,
  {Bartelmann}, M., {Heymans}, C., {Moscardini}, L., \& {Radovich}, M. 2008,
  \aap, 482, 403

\bibitem[{{Pace} {et~al.}(2007){Pace}, {Maturi}, {Meneghetti}, {Bartelmann},
  {Moscardini}, \& {Dolag}}]{2007A&A...471..731P}
{Pace}, F., {Maturi}, M., {Meneghetti}, M., {Bartelmann}, M., {Moscardini}, L.,
  \& {Dolag}, K. 2007, \aap, 471, 731

\bibitem[{{Petkova} {et~al.}(2013){Petkova}, {Metcalf}, \&
  {Giocoli}}]{PM&G2013}
{Petkova}, M., {Metcalf}, R., \& {Giocoli}, C. 2013, ?

\bibitem[{{Poindexter} \& {Kochanek}(2010)}]{2010ApJ...712..658P}
{Poindexter}, S. \& {Kochanek}, C.~S. 2010, \apj, 712, 658

\bibitem[{{Sato} {et~al.}(2009){Sato}, {Hamana}, {Takahashi}, {Takada},
  {Yoshida}, {Matsubara}, \& {Sugiyama}}]{2009ApJ...701..945S}
{Sato}, M., {Hamana}, T., {Takahashi}, R., {Takada}, M., {Yoshida}, N.,
  {Matsubara}, T., \& {Sugiyama}, N. 2009, \apj, 701, 945

\bibitem[{{Schneider} {et~al.}(1992){Schneider}, {Ehlers}, \& {Falco}}]{SEF92}
{Schneider}, P., {Ehlers}, J., \& {Falco}, E.~E. 1992, Gravitational Lenses
  (Springer-Verlag)

\bibitem[{{Takahashi} {et~al.}(2011){Takahashi}, {Oguri}, {Sato}, \&
  {Hamana}}]{2011ApJ...742...15T}
{Takahashi}, R., {Oguri}, M., {Sato}, M., \& {Hamana}, T. 2011, \apj, 742, 15

\bibitem[{{Vale} \& {White}(2003)}]{2003ApJ...592..699V}
{Vale}, C. \& {White}, M. 2003, \apj, 592, 699

\bibitem[{{Vegetti} {et~al.}(2010){Vegetti}, {Koopmans}, {Bolton}, {Treu}, \&
  {Gavazzi}}]{2010MNRAS.tmp.1510V}
{Vegetti}, S., {Koopmans}, L.~V.~E., {Bolton}, A., {Treu}, T., \& {Gavazzi}, R.
  2010, \mnras, 1510

\bibitem[{{Wambsganss} {et~al.}(1998){Wambsganss}, {Cen}, \&
  {Ostriker}}]{1998ApJ...494...29W}
{Wambsganss}, J., {Cen}, R., \& {Ostriker}, J.~P. 1998, \apj, 494, 29

\end{thebibliography}

\appendix

\section{Magnification matrix from deflections}
\label{appendix:magn-matr-from}

In this appendix, it is shown how the magnification matrix can be estimated from three points on the image plane and their corresponding points on the source plane.  This is useful for avoiding the direct calculation of the  lensing quantities with the deflection solver to high precision.

The magnification matrix is the Jacobian matrix of the map between the source and the image planes,
\begin{align}\label{eq:magmatrix}
{\mathcal A} &= \left[ \frac{\partial {\bf y}}{\partial {\bf x}} \right] = \left(
\begin{array}{cc}
1-\kappa - \gamma_1 & -\gamma_2 \\
-\gamma_2 & 1-\kappa + \gamma_1
\end{array}
\right) \\
&= \left(
\begin{array}{cc}
a_{11} & a_{21} \\
a_{12} & a_{22}
\end{array}
\right)
\end{align}
where $\kappa$, $\gamma_1$ and $\gamma_2$ are the lensing quantities referred to in equations~(\ref{kappa}) through (\ref{gamma_2}).  For a single plane lens $a_{12} = a_{21}$ as implied in (\ref{eq:magmatrix}), but in general this may not be the case because of numerical noise or multi-plane lensing (see Paper II) and we will not impose this restriction here.

If we have three points on the image plane (${\bf x}^{(0)}$, ${\bf x}^{(1)}$
and ${\bf x}^{(2)}$) and the corresponding source points (${\bf y}^{(0)}$, ${\bf y}^{(1)}$ and ${\bf y}^{(2)}$) we can define the vectors $\Delta{\bf x}^{(1,2)} = {\bf x}^{(1,2)} - {\bf x}^{(0)}$ and $\Delta{\bf y}^{(1,2)} = {\bf y}^{(1,2)} - {\bf y}^{(0)}$.  To linear order in an expansion of the lensing deflection field $\Delta  {\bf y}^{(1,2)} = {\mathcal A} \Delta  {\bf x}^{(1,2)}$.  These equations can be rewritten in matrix form as
\begin{equation}
\left(
\begin{array}{c}
\Delta y^{(1)}_1 \\
\Delta y^{(1)}_2 \\
\Delta y^{(2)}_1 \\
\Delta y^{(2)}_2 
\end{array}
\right) = 
\left(
\begin{array}{llll}
\Delta x^{(1)}_1 & \Delta x^{(1)}_2 & 0 & 0 \\
0 & 0 & \Delta x^{(1)}_1 & \Delta x^{(1)}_2 \\
\Delta x^{(2)}_1 & \Delta x^{(2)}_2 & 0 & 0 \\
0 & 0 & \Delta x^{(2)}_1 & \Delta x^{(2)}_2 
\end{array}
\right) 
\left(
\begin{array}{c}
a_{11} \\
a_{21} \\
a_{12} \\
a_{22}
\end{array}
\right)
\end{equation}
As long as the three image points are not colinear the matrix can be inverted to solve for the elements of the magnification matrix.  The result is
\begin{align}
a_{11} = \left( x^{(2)}_2 y^{(1)}_1 -  x^{(1)}_2 y^{(2)}_1\right)/{\rm det} \\
a_{22} = \left( x^{(1)}_1 y^{(2)}_2 -  x^{(2)}_1 y^{(1)}_2\right)/{\rm det} \\
a_{12} = \left( x^{(2)}_2 y^{(1)}_2 -  x^{(1)}_2 y^{(2)}_2\right)/{\rm det} \\
a_{12} = \left( x^{(1)}_1 y^{(2)}_1 -  x^{(2)}_1 y^{(1)}_1\right)/{\rm det} 
\end{align}
with
\begin{align}
{\rm det} = x^{(1)}_1 x^{(2)}_2 -  x^{(1)}_2 x^{(2)}_1.
\end{align}
This is clearly very fast to implement in a computer.  

A test for uniform magnification over a grid cell is preformed by taking all pairs of neighbors to that cell that are not colinear with the center of the cell and calculating ${\mathcal A}$ by this method and then comparing each element of ${\mathcal A}$ for each pair of points.  If none of the elements differ by more than the tolerance level the magnification matrix is considered to be uniform over the cell.

\section{singular isothermal sphere}
\label{app:sing-isoth-sphere-1}

Here we summarize some standard analytic results for a Singular
Isothermal Sphere (SIS) lens for reference.

The lensing equation in this case is
\begin{equation}
{\bf y} = {\bf x} - r_e \frac{{\bf x}}{| {\bf x} |}.
\end{equation}
where $r_e$ is the Einstein radius
\begin{equation}\label{eq:einstein_radius}
r_e = 4\pi \left( \frac{\sigma}{c}\right)^2 \frac{\Da_{sl}\Da_l}{\Da_s}
\end{equation}
and $\sigma$ is the velocity dispersion.
For a point source with $|{\bf y}| < r_e$ there are two images.  One has positive
magnification and one has a negative magnification (negative parity).
Their positions are
\begin{equation}
{\bf x}_\pm = {\bf y} \left(  1 \pm \frac{r_e}{| {\bf y} |} \right)
\end{equation}
Only the positive magnification image exists if  $|{\bf y}| > r_e$.
The magnifications of the images are
\begin{equation}\label{eq:sis_point_magnifications+}
\mu_+({\bf y}) = 1 + \frac{r_e}{|{\bf y}|}.
\end{equation}
\begin{equation}\label{eq:sis_point_magnifications-}
\mu_{-}({\bf y}) = \left\{ 
\begin{array}{ccc}
1 - \frac{r_e}{|{\bf y}|}  &,& |{\bf y}| < r_e \\
0 &,& |{\bf y}| > r_e
\end{array}
\right.
\end{equation}

The magnification of a finite size source with surface brightness $f({\bf
  y})$ is
\begin{equation}
\mu_\pm = \frac{1}{F} \int d^2y~ \mu({\bf y}) f({\bf y})
\end{equation}
where $F$ is the integral of $f({\bf y})$ over all ${\bf y}$.  For a
circular, constant surface brightness source this integral can be done
numerically.    When the source intersects with the center of the
lens, the negative and positive images will be joined creating an
Einstein ring.  In the section~\ref{sec:sing-isoth-sphere},  the
joined images are considered to be a single positive magnification
image since this is how GLAMER classifies them.

\end{document}